\documentclass[sigconf]{acmart}
\AtBeginDocument{%
  }

\PassOptionsToPackage{table,xcdraw}{xcolor}
\usepackage{acmart-taps}
\usepackage{pifont}
\usepackage{caption}
\usepackage{subcaption}
\usepackage{listings}
\usepackage{geometry}
\usepackage{booktabs}
\usepackage{multirow}
\usepackage{fvextra}
\usepackage{longtable}
\usepackage{enumitem}
\usepackage[framemethod=TikZ]{mdframed-apt}
\makeatletter
\newcommand{\appentry}[4]{%
  \@dottedtocline{1}{0em}{2.3em}{#1\hspace{0.5em}#2}{#3}%
  \ifx&#4&%
  \else
    \vspace{-2pt}
    \noindent\hspace*{2.3em}\small #4\normalsize
    \vspace{2pt}
  \fi
}
\makeatother

\makeatletter
\newcommand{\appentryy}[4]{%
  {#1\hspace{0.5em}#2}{.........#3}%
  \ifx&#4&%
  \else
    \vspace{-2pt}
    \noindent\hspace*{2.3em}\small #4\normalsize
    \vspace{2pt}
  \fi
}
\makeatother

\newmdenv[
  backgroundcolor=gray!8,
  linecolor=black!30,
  linewidth=0.3pt,
  roundcorner=3pt,
  innerleftmargin=6pt,
  innerrightmargin=6pt,
  innertopmargin=4pt,
  innerbottommargin=4pt
]{quotecallout}
\usepackage{marvosym}

\definecolor{purposegreen}{HTML}{D4EDCF} 
\definecolor{mechyellow}{HTML}{FBF3D0} 
\definecolor{evalpurple}{HTML}{E8DDEF}
\usepackage[table]{xcolor}
\usepackage{colortbl}
\usepackage{tabularx}
\usepackage{float}

\newcommand{\sysname}{\textsc{Sci\-deator}}
\DefineVerbatimEnvironment{Verbatim}{Verbatim}{breaklines=true,breakanywhere=true}

\AtBeginDocument{%
  }

\begin{document}

\acmYear{2026}\copyrightyear{2026}
   \setcopyright{cc}
   \setcctype[4.0]{by}
   \acmConference[ACM CAIS '26]{ACM Conference on AI and Agentic Systems}{May 26--29, 2026}{San Jose, CA, USA}
   \acmBooktitle{ACM Conference on AI and Agentic Systems (ACM CAIS '26), May 26--29, 2026, San Jose, CA, USA}
   \acmDOI{10.1145/3786335.3813161}
   \acmISBN{979-8-4007-2415-2/26/05}

\title{\sysname: Human-LLM Compound System for Scientific Ideation through Facet Recombination and Novelty Evaluation}

\author{Marissa Radensky$^{\text{\textcolor{purple}{\large\ding{96}}}}$}
\authornote{\large{\textcolor{purple}{\ding{96}}\,\textbf{Equal Contributors.}}}
\email{marissaradensky@gmail.com}
\affiliation{%
  \institution{University of Washington}
  \city{Seattle}\state{Washington}\country{USA}
}

\author{Simra Shahid$^{\text{\textcolor{purple}{\large\ding{96}}}}$}
\authornotemark[1]
\email{simra.sshahid@gmail.com}
\affiliation{%
  \institution{Microsoft}
  \city{Delhi}\country{India}
}

\author{Raymond Fok}
\email{rayfok@cs.washington.edu}
\affiliation{%
  \institution{University of Washington}
  \city{Seattle}\state{Washington}\country{USA}
}

\author{Pao Siangliulue}
\email{paos@allenai.org}
\affiliation{%
  \institution{Allen Institute for AI}
  \city{Seattle}\state{Washington}\country{USA}
}
\author{Tom Hope$^{\text{\large$\bigstar$}}$}
\authornote{\large$\bigstar$\,\textbf{Equal Advisors.}}
\email{tomh@allenai.org}
\affiliation{%
  \institution{Allen Institute for AI}
  \city{Seattle}\state{Washington}\country{USA}
}
\author{Daniel S. Weld$^{\text{\large$\bigstar$}}$}
\authornotemark[2]
\email{danw@allenai.org}
\affiliation{%
  \institution{Allen Institute for AI}
  \city{Seattle}\state{Washington}\country{USA}
}
 
\renewcommand{\shortauthors}{Radensky and Shahid {\em et al.}}

\renewcommand{\shortauthors}{Radensky and Shahid {\em et al.}}

\begin{abstract}
The scientific ideation process often involves blending {facets} of existing papers to create new ideas. We contribute \sysname, {the first human-LLM system} for facet-based scientific ideation. Starting from user-provided papers, \sysname\ extracts key facets---purposes, mechanisms, and evaluations---from these and related papers, allowing users to interactively recombine facets to synthesize ideas. 
{\sysname\ is driven by three design choices: (1)~human-in-the-loop facet recombination, in which users select facets from retrieved papers and the system generates ideas by finding analogies across them via the \textit{Faceted Idea Generator} module; (2)~distance-controlled retrieval via the \textit{Analogous Paper Facet Finder} module, which surfaces papers ranging from the same topic to entirely different areas to provide a spectrum of directions; and (3)~facet-based novelty verification via the \textit{Idea Novelty Checker} module, a retrieve-then-rerank pipeline that helps users to evaluate idea originality using facets.} In a user study with computer science researchers, \sysname\ provided significantly more creativity support than a baseline using the same backbone LLM without our facet-based modules, particularly in idea exploration and expressiveness. 
Ablations further show that the facets benefit the novelty checker: facet-based retrieve-then-rerank surfaces more relevant papers than standard retrieval and re-ranking, and a facet-grounded novelty classifier outperforms classifiers that reason over unstructured ideas and papers.

\end{abstract}


\begin{teaserfigure}
  \centering 
  \includegraphics[width=\textwidth]{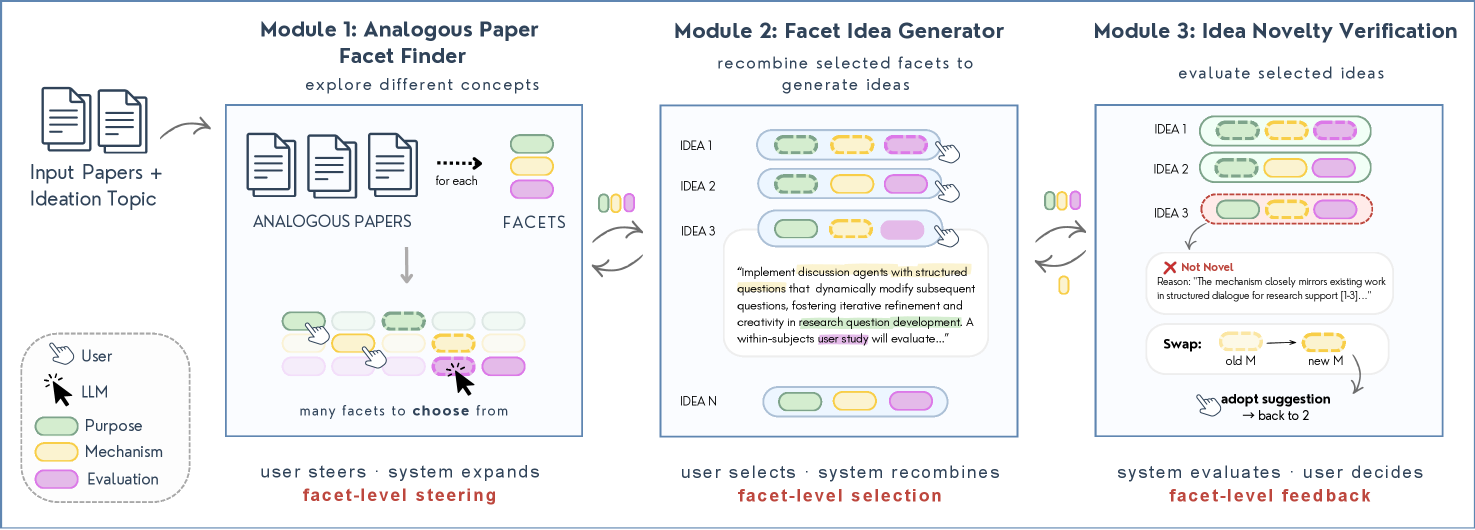}
  \caption{\sysname\ is a human-LLM system for scientific ideation. The user and system interact through idea facets (purposes, mechanisms, evaluations): in Module~1, with the user's input papers and an optional topic, the system retrieves analogous papers and  extracts facets; the user may select interesting facets or leave it to the system. In  Module~2, the system recombines these facets into many ideas via analogy; the user selects ideas to pursue. In Module~3, the system evaluates ideas for novelty against literature retrieved according to facet-based relevance; if  ``not novel,'' it suggests a facet swap. The user can adopt the suggestion, return to Module~2 to regenerate, or return to Module~1 to select or add new facets---an iterative ideation loop.
}
  \label{fig:teaser_actual}
\end{teaserfigure}

\maketitle

\section{Introduction}
Research papers are major sources of scientific inspiration, exposing scientists to concepts that can be recombined into new ideas \cite{portenoy2022bursting,kang2022augmenting,chan2018solvent,Simonton2021ScientificCD}, but discovering inspirations is increasingly challenging due to the ever-expanding literature \cite{bornmann2015growth,jinha2010article} and cognitive fixation that biases scientists toward familiar thinking~\cite{duncker1945problem,purcell1996design}. This challenge has spurred interest in automated systems to assist researchers in discovering new ideas \cite{si2024can,lu2024ai,gottweis2025towards}.
Prior work has highlighted the important role of \emph{faceted} representations for ideation systems: decomposing ideas into their constituent facets has been used to facilitate creative exploration of design spaces and discover abstract structural linkages across ideas \cite{choi2024creativeconnect,suh2024luminate,kang2022augmenting}. Example idea facets include \emph{purposes}, which describe problems addressed, and \emph{mechanisms}, which describe proposed solutions \cite{hope2017accelerating,hope2022scaling}. In the scientific domain, a small body of work has demonstrated the effectiveness of these facets for finding \emph{analogies} between research papers for idea inspiration (e.g., papers with similar purposes using different mechanisms) \cite{chan2018solvent,kang2022augmenting,portenoy2022bursting}.
However, this line of work stopped at surfacing analogous papers as inspirations with no interface for \emph{applying} the inspirations to {synthesize} recombinant ideas or for \emph{evaluating} the generated ideas vis-à-vis existing literature to assess novelty. These important and cognitively taxing tasks were left to the scientists, with no support. 

Meanwhile, large language models (LLMs) raise the prospect of quickly synthesizing and evaluating ideas. Recent work has demonstrated their promise in human-AI interfaces for scientific ideation support \cite{liu2024personaflow,liu2024ai,pu2024ideasynth}. However, none of these human-LLM works explored \textit{facet-based} scientific idea generation, despite the fundamental role facets play in ideation literature. Furthermore, none evaluated interfaces for assessing novelty of generated ideas compared to existing papers.

We present \sysname, the first human-LLM compound system for facet-based scientific idea generation and novelty evaluation.  {\sysname\ composes three LLM-powered modules --- \textit{Analogous Paper Facet Finder} for distance-controlled retrieval of facets, \textit{Faceted Idea Generator} for analogy-based idea generation, and \textit{Idea Novelty Checker} for novelty verification --- into a system where the human directs each stage.} The system uses a faceted representation (purposes, mechanisms, and evaluations) across all modules: the same representation used to describe retrieved papers is also used to compose ideas, rank related work for novelty assessment, and suggest modifications for non-novel ideas. 

The user actions are also at the facet level --- selecting a purpose, swapping a mechanism, adopting a novelty suggestion --- which propagates to each module, giving a structured signal rather than long textual feedback for models to interpret. {Throughout, judgment and idea selection remain with the user: \sysname\ proposes facets, ideas, and novelty assessments, but the user decides which ideas to pursue and which novelty signals to weigh.} With our novel system design and user study, we fill an important knowledge gap regarding the potential of faceted representations for human-AI compound systems for creative tasks such as scientific ideation.

\begin{figure*}[t!]
\centering
\vspace{-3mm}
  \includegraphics[width=\textwidth]{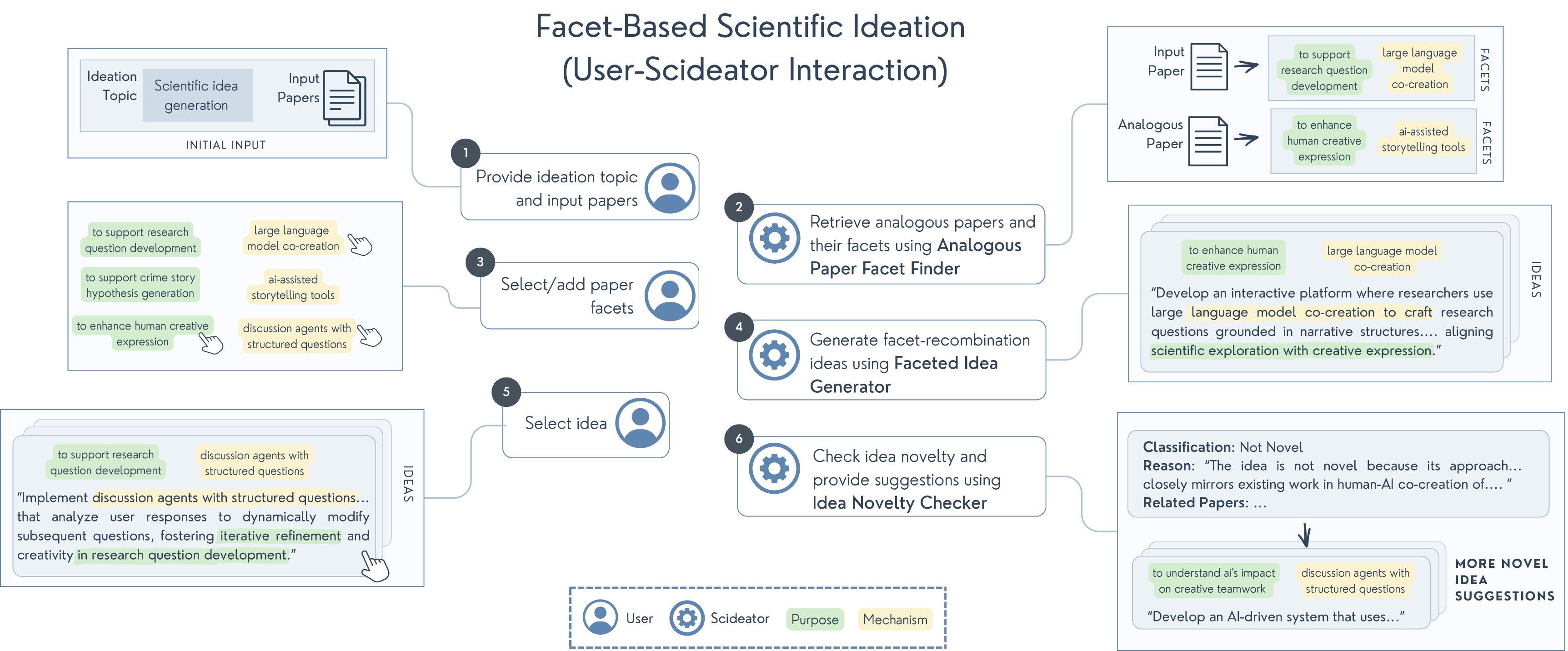}

  \caption{\sysname\ is a human-LLM compound system 
for facet-based scientific ideation.  1)~The user provides an 
ideation topic and input papers. 2)~\sysname\ 
retrieves analogous papers and extracts facets from 
both input and analogous papers. The same 
faceted representation ---\colorbox{purposegreen}{\strut purpose} and \colorbox{mechyellow}{\strut mechanism}---
flows across all stages. The evaluation facets are omitted in the figure for clarity. 3)~The user selects 
facets or adds their own; if no selection is made, 
the system selects automatically. 4)~\sysname\ 
generates ideas by recombining selected facets 
through analogy. 5)~The user selects an idea to 
evaluate. 6)~\sysname\ classifies the idea as 
``novel'' or ``not novel'' with a rationale 
referencing specific papers. The user reviews the 
classification and can adjust it. If ``not 
novel,'' \sysname\ suggests alternatives by swapping 
one facet. The user can adopt a suggestion and 
return to step 3, creating an iterative loop.}

  \Description{There are three columns labeled, from left to right: 1) "User Input [arrow] Tool Output", 2) "User Action" with a human icon, and 3) "Tool Action" with a light bulb icon. In the first row under the column headers, the contents of each cell are: 1) "Papers [arrow] Facets"; 2) "1. Provide ideation topic and input papers.", [image of light blue box labeled "scientific idea generation" and set of blue papers]"; and 3) "2. Retrieve analogous papers and their facets using Analogous Paper Facet Finder.", [image of input paper pointing to input paper purpose (to support research question development) and mechanism (large language model-based agent system) as well as analogous paper pointing to analogous paper purpose (to support crime story hypothesis generation) and mechanism (discussion agents with structured questions)]. In the next row, the cells contain: 1) "Facets [arrow] Ideas"; 2) "3. Select/add paper facets. [Image of three listed purposes-- to support research question development, to support crime story hypothesis generation, and to recommend biomedical research directions, followed by an ellipsis. The first purpose is highlighted. Next to the purposes is a list of three mechanisms-- large language model-based agent system, discussion agents with structured questions, and deep learning ranking criteria, followed by an ellipsis.]; and 3) "4. Generate facet-recombination ideas using Faceted Idea Generator.", [image of two gray boxes that each contain a purpose and a mechanism. Both have the purpose 'to support research question development.' One has the mechanism 'discussion agents with structured questions' and the other 'deep-learning ranking criteria.']. In the next row, the cells contain: 1) "Idea [arrow] Novelty Check"; 2) "5. Select idea.", [image same as in last cell described but first gray box is highlighted]; and 3) "6. Check idea novelty and provide more novel idea suggestions (if not novel) using Idea Novelty Checker.", [image of first gray box with purpose and mechanism from last cell, no longer highlighted, a box underneath that says "Related Papers: ..., Classification: Not Novel, Reason: This idea is not novel because...", and beneath that the label "more novel idea suggestions" with a curly brace encircling two gray boxes with a purpose and mechanism. The first gray box says "to recommend biomedical research directions" in italics/bold for purpose but otherwise has the same mechanism as the current idea. The second gray box says "human-in-the-loop knowledge graph" in italics/bold for mechanism but otherwise has the same purpose as the current idea.]}
  \label{fig:teaser}
\end{figure*}

To use \sysname\ (Fig \ref{fig:teaser}), scientists provide input papers and an optional ideation topic. \sysname\ extracts key facets (purpose, mechanism, and evaluation) from the input papers; the evaluation facet describes the paper's method to determine if the mechanism successfully addressed the purpose. \sysname\ retrieves papers with purpose-mechanism pairs analogous to the input papers' overarching purpose and mechanism. Scientists work with \sysname\ to select candidate facets from retrieved and input papers for recombination. The tool generates analogies involving candidate facets and produces ideas based on the most promising ones.

\sysname\ also provides idea novelty classifications with explanations, retrieving relevant literature and using carefully constructed in-context examples labeled by experts. The classifier helps researchers make informed judgments by showing facet-level overlap of generated idea and retrieved papers. {Building on combinatorial creativity~\cite{boden2004creative,Simonton2021ScientificCD} and facet-based analogy in scientific ideation~\cite{hope2017accelerating,hope2022scaling}, we define an idea as novel if it differs from retrieved papers in at least one core facet, uniquely combines these facets, or applies them to a new domain.} The system also provides suggestions for improving the novelty of ideas, by replacing one of the initial idea's facets to make it novel.

We investigate the impact of facet-based interaction on scientific ideation through a within-subjects user study with 22 computer-science researchers. Participants completed ideation sessions with two tools: \sysname\ and a baseline that uses the same backbone LLM without our facet-based modules. We analyze how \sysname\ impacts aspects of the user's creative experience, such as the idea exploration process and the perceived ability to express oneself.
Participants experienced significantly more creativity support with \sysname, particularly in exploring different ideas, which they considered the most important factor for creativity support.
Importantly, when participants discussed finding new concepts with \sysname, they cited the tool's facets or ideas as the source, whereas with the baseline, they most often cited the input papers as the source. This shows that \sysname\ extends users' thinking beyond initial perspectives (represented by input papers), while a similar interface without a facet-based interaction tends to provide ideas only within those confines.

Results also suggest \sysname's novelty checker helps to filter unoriginal ideas.  Participants tended to lower their idea novelty  assessments  when \sysname\ classified the idea as `not novel,' which they could verify using provided related papers and explanations.
{We also conduct ablations of the retrieval and classification components of the novelty checker that further show that the faceted representation benefits both. Holding the classifier constant, facet-based re-ranking ranks relevant prior work higher than general-relevance re-ranking or embedding-only retrieval (Sec \ref{sec:retrieval_ablation_results}). Keeping retrieval constant, the facet-grounded classification outperforms classifiers that reason over ideas and papers as wholes (Sec \ref{sec:classifier_ablation}).} Together with the user study findings, these results provide the first systematic evaluation of novelty assessment within a human-AI ideation system---examining both the retrieval and classification pipeline and how researchers use novelty judgments in practice.

\vspace{1mm}
\noindent\textbf{In summary, we make the following contributions:}

\begin{itemize}[leftmargin=*,topsep=0pt]

\item \sysname, the first human-LLM compound system for scientific ideation that maintains a single faceted representation (purposes, mechanisms, and evaluations) across 
retrieval, generation, novelty evaluation and suggestion, where users can select, modify, or build on the system's outputs at each stage.

\item  A within-subjects user study (N=22) showing that \sysname\ provides significantly more creativity support than a baseline using the same LLM without faceted interaction, with evidence that participants discovered new concepts through the system's facets and preferred facet-level steering over prompting.

\item {Evaluation of our novelty checker, with a retrieval ablation showing our facet-based re-ranking improves paper selection for novelty assessment, and a classifier ablation showing the facet-grounded classification independently outperforms baseline novelty checkers on the same retrieved papers.} Combined with user study evidence of how researchers interpret and  act on novelty assessments, this provides the first systematic evaluation of  novelty assessment within a human-AI ideation system.

\end{itemize}

\noindent 
The remainder of the paper is organized as follows.  We describe \sysname's building blocks and shared faceted representation (Section~\ref{sec:building_blocks}),  its end-to-end workflow (Section~\ref{sec:workflow}),  and experiments evaluating the novelty checker and  the user study (Section~\ref{sec:experiments}),  followed by related work  (Section~\ref{sec:relatedwork}) and discussion  (Section~\ref{sec:discussion}). We encourage readers to consult the appendix for the full user study design along with UI screenshots, implementation details, LLM prompts used across all modules, and sample ideas generated by the system.

\section{Building Blocks of \sysname}
\label{sec:building_blocks}
In this section we describe the individual modules that make up \sysname. We then describe the end-to-end workflow in Section \ref{sec:workflow}, illustrating how these components of facets, analogous facet generation and paper retrieval, idea generation, and novelty assessment work together in a continuous, human-directed ideation loop.

\subsection*{Shared Representation: Paper Facets}

\sysname\ represents ideas and papers using three facets: the purpose (the problem being addressed), the mechanism (the proposed solution), and the evaluation (the method for determining whether the solution works). We add the evaluation facet to the purpose and mechanism facets used in prior work \cite{hope2017accelerating,hope2022scaling} because evaluation is an important part of a research idea. Facets are extracted by prompting an LLM to produce short phrases (no more than 7 words) based on a paper's title and abstract.

This faceted representation is maintained across 
every stage of the pipeline. The purpose and mechanism facets
drive retrieval --- the system finds analogous papers 
by matching these two facets --- while all three 
facets are used to describe retrieved papers, compose 
ideas, assess novelty, and suggest improvements. For the user, this means working with a single representation throughout; rather than switching between free-text prompts and paper abstracts, they select, combine, and revise the same structured facets at each stage.  For the system, these user interactions at the facet level become precise signals that directly shape what the next module produces.

Next we describe the three modules that leverage this shared representation for different stages of the ideation pipeline. 

\vspace{-2mm}

\subsection*{Module 1: Analogous Paper Facet Finder}\label{sec:anlgretrival}

This module takes a set of papers and retrieves 
analogous papers based on their purpose-mechanism 
pairs at controlled conceptual distances, then 
extracts all three facets (including evaluation) 
from the retrieved papers.

Consider the example in Figure~\ref{fig:teaser}: the input paper has purpose \colorbox{purposegreen}{support research question development} and mechanism \colorbox{mechyellow}{large} \colorbox{mechyellow}{language model co-creation}. The module first extracts this purpose and mechanism, then generates analogous purpose-mechanism pairs at three distance levels:
\begin{itemize}
\item \textbf{Near} (same topic): pairs that address a similar problem with a different approach within the same research topic. For instance, consider the pair (purpose: \colorbox{purposegreen}{support crime story} \colorbox{purposegreen}{hypothesis generation} and the mechanism: \colorbox{mechyellow}{discussion} \colorbox{mechyellow}{agents with structured questions}). Both this and the input paper's pair involve guiding a user through structured dialogue to develop ideas, but in a different application context.
\item \textbf{Far} (same subarea, different topic): pairs from a related area of computer science that share a more abstract structural parallel with the input. A far analogous paper might have the purpose \colorbox{purposegreen}{enhance human creative expression} and mechanism \colorbox{mechyellow}{ai-assisted storytelling tools}, still within human-AI collaboration but addressing a different topic (creative expression rather than pure idea development).
\item \textbf{Very far} (different subarea entirely): pairs from a different area of computer science, connected only by a high-level analogy to the input.
\end{itemize}

The module generates four facet-pairs at each distance level, giving the user a range of options to explore at each level of similarity. Each of these 12 pairs (3 distances x 4 facet pairs) is then grounded in the literature. For each pair, the LLM generates a search query, and the module retrieves matching papers from Semantic Scholar, shortening the query iteratively if no results are found. The first retrieved paper becomes the representative for that pair and the remaining three provide additional context. Separately, four \textbf{very near} papers (most similar to the input) are retrieved directly via Semantic Scholar's paper similarity endpoint~\cite{Kinney2023TheSS}. The LLM then extracts purpose, mechanism, and evaluation facets from all retrieved papers (16 total across four distance levels)\footnote{A small-scale annotation study assessing the consistency of the system's distance labels is reported in 
Appendix~\ref{app:module_annotation}.}. The module also produces a summary of relevant works from the input and very-near papers, which the idea generator (Section~\ref{sec:ideagen}) uses to differentiate new ideas from existing work.

\subsection*{Module 2: Faceted Idea Generator}\label{sec:ideagen}
This module takes a pool of facets from input and retrieved papers and generates ideas by recombining them via analogy, in three steps.  
First, for two sets of papers from different distance groups (e.g., input, near), the LLM generates six candidate analogies between their purpose-mechanism pairs. Continuing the 
earlier example, one analogy between the input paper, with the purpose \colorbox{purposegreen}{support research question development}, and the near paper, with the mechanism \colorbox{mechyellow}{discussion agents with structured questions}, might highlight that both involve using structured dialogue to iteratively guide a user toward generating new ideas. Second, the LLM selects the two strongest analogies based on idea quality (understandability, relevance, feasibility, specificity, and novelty). Third, each selected analogy is converted into an idea: one idea combines a purpose from the first paper set with a mechanism from the second paper set, and the other does the reverse (Figure~\ref{fig:teaser}, steps 3-4).

The module is designed to combine papers from different distances to produce a range of ideas. When combining with near papers, ideas tend to stay within the same research area but apply a different approach. When combining with far or very far papers, ideas cross domain boundaries. The module adapts to user 
selections: (i)~when \textbf{no facets are selected}, the user is 
likely exploring, so the module combines facets from input and 
very-near papers with facets from near, far, and very-far papers 
to maximize diversity; (ii)~when only a \textbf{purpose or 
mechanism is selected}, the user has a direction in mind but wants 
the system to fill in the rest, so the module pairs the selected 
facet with complementary facets from papers at different distances; 
and (iii)~when \textbf{both are selected}, the user has a specific 
combination in mind, and the module combines them directly. In all 
cases, the module differentiates ideas from existing work described 
in the relevant-works summary and self-critiques against quality 
criteria before finalizing each idea.

\subsection*{Module 3: Idea Novelty Verification}\label{sec:ideanovelty_module}

This module takes an idea and determines whether it is novel relative to existing literature. It operates through a four-step retrieve-then-rerank pipeline (Figure~\ref{fig:noveltyChecker-workflow} in Appendix): 1) retrieve candidate relevant papers, 2) select most relevant papers, 3) evaluate idea novelty, and 4) suggest more novel ideas.

\vspace{2mm}

\textbf{\textsc{Step 1: Retrieve candidate relevant papers.}}
The module assembles a broad collection of papers that might overlap with the idea. This includes all papers from prior modules and their related papers via the Semantic Scholar API~\cite{Kinney2023TheSS}. However, simple retrieval methods often overlook contextual aspects of ideas such as their purpose, mechanism and evaluation facets \cite{mysorecsfcube, mysore2022multi, wang2023doris} when assessing similarity between ideas-papers. To expand the coverage of papers, the module generates keyword-based search queries from the idea (LLM-extracted keywords and potential titles) and retrieves matching papers, a popularly used query-based retrieval method also used in~\cite{lu2024ai, si2024can}. Because keyword searches can introduce irrelevant results, the module complements them with Semantic Scholar's snippet-text search\footnote{\href{https://api.semanticscholar.org/api-docs/\#tag/Snippet-Text}{api.semanticscholar.org/api-docs/\#tag/Snippet-Text}}, which matches the full idea text against passages from papers to identify contextually relevant work that keyword queries may miss.

\textbf{\textsc{Step 2: Select most relevant papers.}}
To surface papers most likely to overlap with the idea, the module applies a two-stage re-ranking process following established retrieve-then-rerank practices~\cite{Gao2024LLMenhancedRI, Nouriinanloo2024ReRankingSB, Abdallah2025RankifyAC, Meng2024RankedLT, Sun2023IsCG, Baldelli2024TWOLARAT}. The first stage uses \textbf{embedding-based filtering}: SPECTER embeddings~\cite{Cohan2020SPECTERDR} compute semantic similarity between the idea and each candidate paper, selecting the top $N$ (default 100). This embedding-based ranking efficiently narrows down the paper collection but, compared to LLMs~\cite{Reimers2019SentenceBERTSE}, fails to capture more contextual relationships between different facets of the idea and related papers. The second stage applies \textbf{facet-based LLM re-ranking} using RankGPT~\cite{sun2023chatgpt}. Unlike general relevance re-ranking, this step compares each paper against the idea's specific application domain, purpose, mechanism, and evaluation. This facet-based ranking reflects our novelty definition (Step 3): an idea is not novel if its core facets overlap with existing work, so papers with more facet overlap are most relevant for assessment. Papers are ranked by decreasing facet overlap: (1) papers matching all key facets, (2) papers matching application domain and purpose, (3) papers matching purpose, mechanism, or evaluation, and (4) papers with partially matched or related facets. The top $k$ papers (default 10) proceed to novelty assessment.

\textbf{\textsc{Step 3: Evaluate idea novelty.}}
Using the top-$k$ papers, the module prompts an LLM  to classify the idea as ``novel'' or ``not novel,''  accompanied by reasoning that references specific  papers. The prompt includes expert-labeled in-context examples, each comprising an idea, top-$k$ papers, a novelty label, and a classification reason (see examples, Appendix~\ref{app:expert_examples}). {Examples encode our novelty definition, grounded in combinatorial creativity~\cite{boden2004creative,Simonton2021ScientificCD} and facet-based analogy in scientific ideation~\cite{hope2017accelerating,hope2022scaling}. An idea is novel if it (1) differs from retrieved papers in at least one core facet (a facet-level delta~\cite{hope2017accelerating,hope2022scaling}), (2) uniquely combines facets (a new combination~\cite{boden2004creative}), or (3) applies them to a new domain (analogical transfer~\cite{holyoak1996mental}).}

\textbf{\textsc{Step 4: Suggest more novel ideas.}}
When an idea is classified as ``not novel,'' the module generates three suggestions (one per facet 
type) for more novel alternatives. Each suggestion replaces a different facet in the original idea (purpose, mechanism, or evaluation) with another available facet, aiming to increase novelty relative to the retrieved papers. 
\paragraph{\textbf{Expert Annotation Study.}} The  in-context examples and novelty definition above  emerged from an annotation study in which the first  two authors assessed the novelty of 51 ideas (46  generated by \sysname, 5 adapted from OpenReview)  based on top-10 retrieved papers. An initial round using three categories  (novel, moderately novel, not novel) achieved moderate agreement   (Cohen's $\kappa$ = 0.64), with disagreements  arising partly because annotators drew on background  knowledge rather than the retrieved papers. This led  us to simplify the scheme to binary (novel / not  novel), require judgments based solely on retrieved  papers, and adopt the facet-based definition used  throughout the novelty checker. The revised round  yielded higher agreement (Cohen's $\kappa$ = 0.68).  From both rounds, we collected 67 consensus-labeled examples (39 novel, 28 not novel). We split these into a training pool, from which we sample in-context examples,\footnote{We tested with 10, 15, and 20 examples per class. The module uses 20 examples per class to assess novelty.} and a held-out test set, used for the ablation studies in Section~\ref{sec:retrieval_ablation}.

\paragraph{\textbf{Implementation details}} All model choices, API 
parameters, and default settings are in 
Appendix~\ref{app:implementation}; LLM prompts in~\ref{app:prompts}.

\section{End-to-End Workflow}\label{sec:workflow}

The modules above compose into \sysname's ideation loop in which the system proposes and the scientist steers (Figure~\ref{fig:teaser}). A scientist begins by providing an ideation topic and one or more input papers. \sysname\ runs the retrieval module, extracting facets from the input papers and retrieving analogous papers at four conceptual distances. The system presents all extracted facets organized by distance (Figure~\ref{fig:treatmentideation} in Appendix). 

Because facets are short phrases (up to 7 words), their meaning may not always be immediately clear, especially for facets from distant domains. To support interpretation, each facet is linked to its source paper, so users can trace where a purpose or mechanism originated. Hovering over a facet reveals a longer description. Users can also type in their own facets or request additional facets with an optional guiding query, which triggers a new retrieval in the direction specified by the query.

The scientist then selects facets and triggers idea generation---or generates without selecting, in which case the system chooses facets. The generator adapts to these selections as described in Section~\ref{sec:ideagen}: the specific combination of facets determines which papers and distances the module draws from. Each generated idea is displayed with its constituent facets color-coded by type (Figure~\ref{fig:treatmentideation} in Appendix). Users can click \textit{Expand} on any idea to see a more detailed version, or add their own idea, which the system decomposes into facets and adds to the available pool for future rounds.

To evaluate an idea's novelty, the scientist opens the novelty checker (Figure~\ref{fig:treatmentevaluation} in Appendix). The system retrieves and re-ranks related papers, classifies the idea, and if its ``not novel'', it suggests three alternatives, each replacing a different facet. The scientist can adjust the classification based on their own judgment, adopt a suggestion, or return to facet selection with new perspective.

This creates an \textbf{iterative loop}. A scientist might generate ideas, check one for novelty, learn it overlaps with existing work, adopt a suggestion that swaps the mechanism, and regenerate --- each step expressed at the facet level. Facets selected in one round may be discarded after evaluation; suggestions from the novelty checker introduce facets the scientist had not considered; user-added facets expand the pool for future rounds.

A key aspect of this design is that every interaction--- selecting a facet, adopting a suggestion, adding a custom query --- gives the system a structured signal rather than free-text instructions.  When the novelty checker classifies an idea as ``not novel'' because its mechanism overlaps with a paper, and the user adopts a suggestion that replaces that mechanism, the system receives a precise signal about which dimension to change. This contrasts with a text-based LLM interaction where the user writes ``make the idea more novel'' and the model must interpret both intent and scope. In \sysname, the faceted representation makes intent explicit and the system's response predictable.

\section{Experiments}\label{sec:experiments}
We evaluate \sysname\ in two parts. Our primary evaluation of system utility is a within-subjects study with 22 computer-science researchers (Section~\ref{sec:userstudy}) that investigates whether the faceted interaction helps researchers generate and evaluate scientific ideas. We also conduct complementary ablation studies of the novelty checker module (Section~\ref{sec:retrieval_ablation}) to examine whether the faceted representation improves the system's ability to retrieve relevant papers and classify novelty.

\subsection{User Study}
\label{sec:userstudy}

To assess whether \sysname's faceted interaction supports scientific ideation in practice, we conducted a within-subjects study comparing \sysname\ to a baseline tool without faceted representation. 22 computer-science researchers participated (16 PhD students, 5 master's students, 1 industry), each completing two 20-minute ideation sessions --- one with \sysname\ and one with the baseline --- in randomized order, using ideation topics and starting papers that we assigned from their respective domains of HCI (12) or NLP (10).\footnote{Three participants focused on other ML areas but were comfortable ideating on NLP topics.} All of the participants had authored at least one paper.

The baseline represents \textit{paper-level interaction}: participants select which of the starting papers the LLM should utilize and provide any free-text instructions, and the model generates ideas directly. Both tools use the same LLM, so the comparison isolates the effect of organizing interaction around a structured faceted representation versus free-text prompting. After each session, participants rated their experience using the Creativity Support Index (CSI)~\cite{cherry2014quantifying}, a validated instrument that measures factors of creative support such as exploration, expressiveness, enjoyment. We also collected interaction logs, survey responses and conducted semi-structured interviews analyzed through inductive thematic analysis~\cite{braun2006using}. Full study design, baseline tool description, procedure, and participant demographics are in Appendix~\ref{app:studydesign}.

We address three research questions. RQ1 tests whether faceted interaction leads to more creativity support. RQ2 examines how participants use facet-level versus paper-level interaction. RQ3 tests whether the novelty checker influences participants' assessments. Sample ideas generated by 
participants using both \sysname\ and the baseline 
are provided in Appendix~\ref{sec:longIdeas}.

\subsubsection{\textbf{RQ1: Does Faceted Interaction Lead to More Creativity Support for Idea Generation?}}\label{sec:rq1}

Participants experienced significantly more creativity 
support with \sysname\ than the baseline (Wilcoxon 
signed-rank test, p<.01; CSI scores on a 0--100 scale: 
\sysname\ median=70.5, baseline median=61.0)\footnote{CSI scores: \sysname\ median=70.50 
(Q1=57.50, Q3=79.00), baseline median=61.00 
(Q1=42.25, Q3=71.50). Wilcoxon signed-rank test, 
V=208.50, p<.01. The data did not violate the 
assumption of symmetry of within-subjects differences 
about the median.}.
Of the CSI factors, participants benefited most from \sysname\ in \textbf{exploration} and \textbf{expressiveness} --- the factors participants ranked as first- and third-most important for this ideation task (Figure~\ref{fig:all_CSI}). Results for factors such as immersion, showed no differences and are reported in Appendix~\ref{app:csi_factors}.

\vspace{1mm}
\noindent \textbf{Exploration Factor.} \sysname\ helped participants explore ideas beyond their input papers. While participants rated their favorite ideas' newness on a 
7-point scale, showing only a slight advantage for 
\sysname\ (median difference: 0.5 points), the qualitative evidence was striking: in the baseline, participants who found new concepts attributed them to the \textit{input papers} rather than the tool's output (6 of 8), while in the treatment, all participants who found new concepts cited the \textit{tool's facets or generated ideas} as the source (6 of 6). For example, P18-HCI-treatment shared, 

\aptLtoX{\begin{mdframed1}``\textit{When I thought of human-AI collaboration in art, for example, I did not think about also supporting artistic pursuits of students [which surfaced in a generated idea].... When I was thinking about the topic, I thought more about... a human prompting an AI for generating images or for image exploration which is more related to the papers that were given.}''\end{mdframed1}}{\begin{mdframed}[
  backgroundcolor=gray!10,
  leftline=true,
  rightline=false,
  topline=false,
  bottomline=false,
  linecolor=black!40,
  linewidth=2pt,
  innerleftmargin=6pt,
  innerrightmargin=4pt,
  innertopmargin=3pt,
  innerbottommargin=3pt, 
  skipabove=10pt,
  skipbelow=10pt
]
``\textit{When I thought of human-AI collaboration in art, for example, I did not think about also supporting artistic pursuits of students [which surfaced in a generated idea].... When I was thinking about the topic, I thought more about... a human prompting an AI for generating images or for image exploration which is more related to the papers that were given.}'' 
\end{mdframed}}
Meanwhile, P5-NLP-baseline reflected, 

\aptLtoX{\begin{mdframed1}``\textit{The papers themselves were really interesting, but I don't think the tool generated anything super beyond a synthesis of the ideas that were in those three papers.}''\end{mdframed1}}{\begin{mdframed}[
  backgroundcolor=gray!10,
  leftline=true,
  rightline=false,
  topline=false,
  bottomline=false,
  linecolor=black!40,
  linewidth=2pt,
  innerleftmargin=6pt,
  innerrightmargin=4pt,
  innertopmargin=3pt,
  innerbottommargin=3pt, 
  skipabove=10pt,
  skipbelow=10pt
]
``\textit{The papers themselves were really interesting, but I don't think the tool generated anything super beyond a synthesis of the ideas that were in those three papers.}''
\end{mdframed}}

This pattern suggests the faceted representation served as a medium for new concepts to reach the user. The retrieval module surfaced papers the user had not started with, the extraction module decomposed them into the same facet format the user was already working with, and participants encountered new concepts through these facets. Four participants also noted that the facet-level interaction supported exploration through \textbf{greater transparency} as they could trace which facets contributed to each idea and more easily context-switch between research directions. For instance, P5-NLP-treatment reflected, 
\aptLtoX{\begin{mdframed1}
``\textit{I think the first thing that I noticed was that it was very easy to context switch. That was my main problem with the [other] tool before. I couldn't figure out which idea dealt with what aspect of the research that I was engaging with. Very easy to do that here.}''
\end{mdframed1}}{\begin{mdframed}[
  backgroundcolor=gray!10,
  leftline=true,
  rightline=false,
  topline=false,
  bottomline=false,
  linecolor=black!40,
  linewidth=2pt,
  innerleftmargin=6pt,
  innerrightmargin=4pt,
  innertopmargin=3pt,
  innerbottommargin=3pt, 
  skipabove=10pt,
  skipbelow=10pt
]
``\textit{I think the first thing that I noticed was that it was very easy to context switch. That was my main problem with the [other] tool before. I couldn't figure out which idea dealt with what aspect of the research that I was engaging with. Very easy to do that here.}''
\end{mdframed}}

\vspace{1mm}
\noindent \textbf{Expressiveness Factor}. Fourteen participants found \sysname's facet-level interaction useful or interesting, with seven specifically noting \textbf{increased control} over idea generation. P11-NLP-treatment explained, 
\aptLtoX{\begin{mdframed1}``\textit{I like this tool better because it sort of distilled the different aspects of the input papers into very concrete blocks that you could plug into each other.... it's just that the information was presented in this tool... in a more digestible manner, and that helped combine information across papers better.}''\end{mdframed1}}{\begin{mdframed}[
  backgroundcolor=gray!10,
  leftline=true,
  rightline=false,
  topline=false,
  bottomline=false,
  linecolor=black!40,
  linewidth=2pt,
  innerleftmargin=6pt,
  innerrightmargin=4pt,
  innertopmargin=3pt,
  innerbottommargin=3pt, 
  skipabove=10pt,
  skipbelow=10pt
]
``\textit{I like this tool better because it sort of distilled the different aspects of the input papers into very concrete blocks that you could plug into each other.... it's just that the information was presented in this tool... in a more digestible manner, and that helped combine information across papers better.}''
\end{mdframed}}
This interest in structured interaction is also reflected in participants' use of custom instructions. With \sysname, participants most often did not provide custom text instructions for generating their 2 favorite ideas (median: 0 of 2). With the baseline, the median participant used custom instructions for 1.5 of 2 favorite ideas. As P1-HCI-treatment put it: 
\aptLtoX{\begin{mdframed1}``\textit{I didn't need to add any custom instructions because these [facets] served like custom instructions.}''\end{mdframed1}}{\begin{mdframed}[
  backgroundcolor=gray!10,
  leftline=true,
  rightline=false,
  topline=false,
  bottomline=false,
  linecolor=black!40,
  linewidth=2pt,
  innerleftmargin=6pt,
  innerrightmargin=4pt,
  innertopmargin=3pt,
  innerbottommargin=3pt, 
  skipabove=10pt,
  skipbelow=10pt
]
``\textit{I didn't need to add any custom instructions because these [facets] served like custom instructions.}''
\end{mdframed}}
The faceted interface provided a structured alternative to free-text prompting that participants found useful. 

\subsubsection{\textbf{RQ2: How Do Participants Use Facet-Level versus Paper-Level Interaction?}}\label{sec:rq2}

Participants commented on benefits of both the baseline tool and \sysname\ in terms of their input granularity.
Fourteen participants found \sysname's affordance for facet-level interactions useful or interesting, noting \underline{increased control} and \underline{greater transparency} (Section \ref{sec:rq1}) as advantages of the facet-level interaction. On the other hand, five
participants appreciated the baseline's paper-level interactions in addition to or more than \sysname's facet-level interactions. Three of these participants liked the paper-level interaction, as it felt more directly connected to the literature. P22-NLP-baseline explained, 
\aptLtoX{\begin{mdframed1}``\textit{ I think papers for me were more natural than facets.... I think to me it's more like a map of literature, so I could see it more with papers.}.''\end{mdframed1}}{\begin{mdframed}[
  backgroundcolor=gray!10,
  leftline=true,
  rightline=false,
  topline=false,
  bottomline=false,
  linecolor=black!40,
  linewidth=2pt,
  innerleftmargin=6pt,
  innerrightmargin=4pt,
  innertopmargin=3pt,
  innerbottommargin=3pt, 
  skipabove=10pt,
  skipbelow=10pt
]
``\textit{ I think papers for me were more natural than facets.... I think to me it's more like a map of literature, so I could see it more with papers.}.''
\end{mdframed}}

\begin{figure}[tb!]
  \includegraphics[width=\columnwidth]{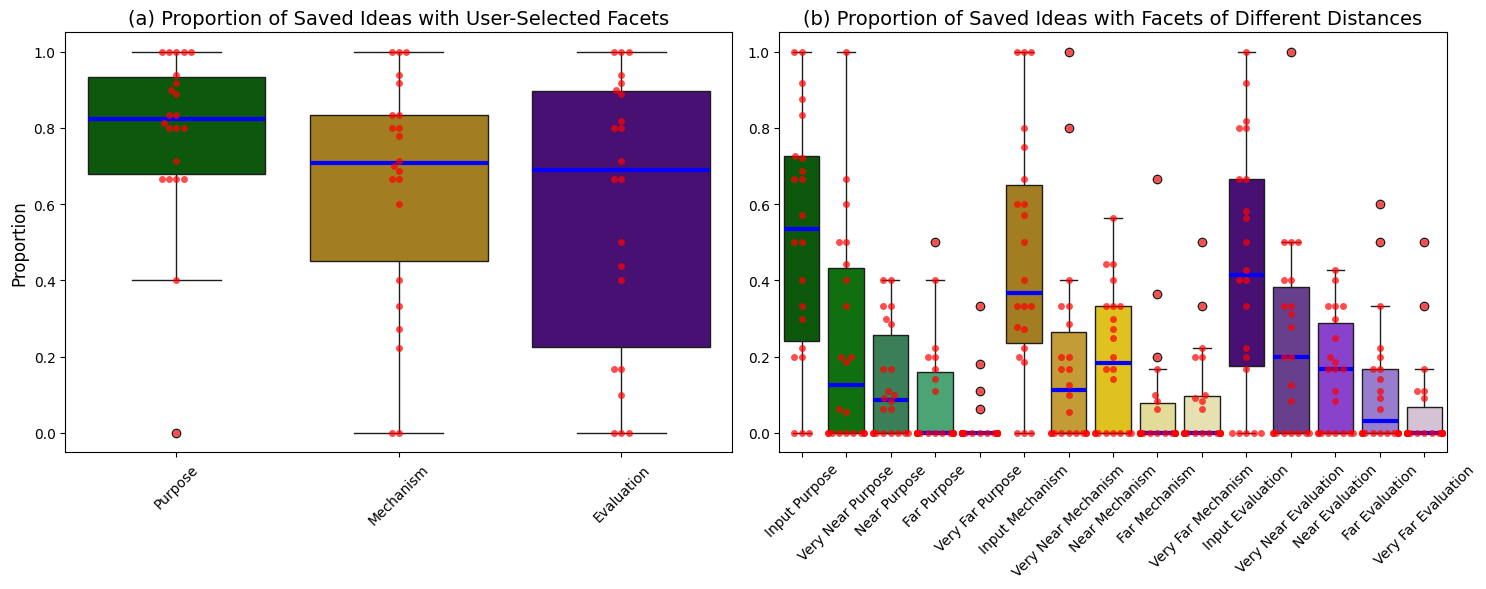}
  \centering
  \caption{(a) Participants more often opted to select their own facets rather than let the LLM select for them. (b) Participants used input facets and facets nearer to the input more than facets farther from the input. }
  \Description{(a) A set of three boxplots titled "Proportion of Saved Ideas with User-Selected Facets." The y-axis is labeled "Proportion" and has ticks from 0.0 to 1.0. The x-axis has ticks for the three boxplots: purpose, mechanism, and evaluation. We see that purpose was most consistently user-selected (M \~ 0.825, Q1 \~ 0.7, Q3 \~ 0.925). Mechanism was next most (M \~ 0.7, Q1 \~ 0.45, Q3 \~ 0.8). Evauation was third most, but still more likely to be user-selected (M \~ 0.7, Q1 \~ 0.2, Q3 \~ 0.9). (b) A set of 15 boxplots titled "Proportion of Saved Ideas with Facets of Different Distances." The y-axis is also labeled "Proportion" with ticks for 0.0 to 1.0. The x-axis has ticks for each boxplot: Input Purpose, Very Near Purpose, Near Purpose, Far Purpose, Very Far Purpose, [the same 5 distances for mechanism], [the same 5 distances for evaluation]. The boxplots indicate that, for the most part, facets were used less and less as their distance increased. Furthermore, there was a particularly strong bias away from farther facets and toward input facets for purpose.}
  \label{fig:facetAspects}
   \vspace{-5mm}
\end{figure}

\noindent \textbf{Facet Selection.} When generating ideas in \sysname, users have the flexibility to either manually curate facet combinations (using system-extracted or user-provided facets) or rely on automated system selection. Participants' favorite ideas more often included evaluations, mechanisms, and especially purposes selected by themselves rather than the LLM (Figure \ref{fig:facetAspects}a). Given that participants were assigned ideation topics, P18-HCI-treatment explained why participants may have decided to prioritize selecting purposes themselves: 
\aptLtoX{\begin{mdframed1}``\textit{I think the purpose is the most relevant to the topic. So within an area, there can be many ways of doing the same tasks, but the task is ultimately what defines the area.}''\end{mdframed1}}{\begin{mdframed}[
  backgroundcolor=gray!10,
  leftline=true,
  rightline=false,
  topline=false,
  bottomline=false,
  linecolor=black!40,
  linewidth=2pt,
  innerleftmargin=6pt,
  innerrightmargin=4pt,
  innertopmargin=3pt,
  innerbottommargin=3pt, 
  skipabove=10pt,
  skipbelow=10pt
]
``\textit{I think the purpose is the most relevant to the topic. So within an area, there can be many ways of doing the same tasks, but the task is ultimately what defines the area.}''
\end{mdframed}}

\vspace{1mm}
\noindent\textbf{Distance Preferences.} Participants used input and near facets substantially more than far facets, primarily for purposes (Figure \ref{fig:facetAspects}b). The most common reason for not using far facets was lower perceived relevance to the ideation topic, though four participants found them helpful for discovery. P9-NLP-treatment commented, 
\aptLtoX{\begin{mdframed1}``\textit{that very near, near, far kind of thing... it kind of adds some sort of discovery factor.}''\end{mdframed1}}{\begin{mdframed}[
  backgroundcolor=gray!10,
  leftline=true,
  rightline=false,
  topline=false,
  bottomline=false,
  linecolor=black!40,
  linewidth=2pt,
  innerleftmargin=6pt,
  innerrightmargin=4pt,
  innertopmargin=3pt,
  innerbottommargin=3pt, 
  skipabove=10pt,
  skipbelow=10pt
]
``\textit{that very near, near, far kind of thing... it kind of adds some sort of discovery factor.}''
\end{mdframed}}

\subsubsection{\textbf{RQ3: Does the Idea Novelty Verification Module Influence Participants' Confidence in Novelty Assessments?}}\label{sec:rq3}

\begin{figure}[tb]
  \centering
  \vspace{-3mm}
  \includegraphics[width=\columnwidth]{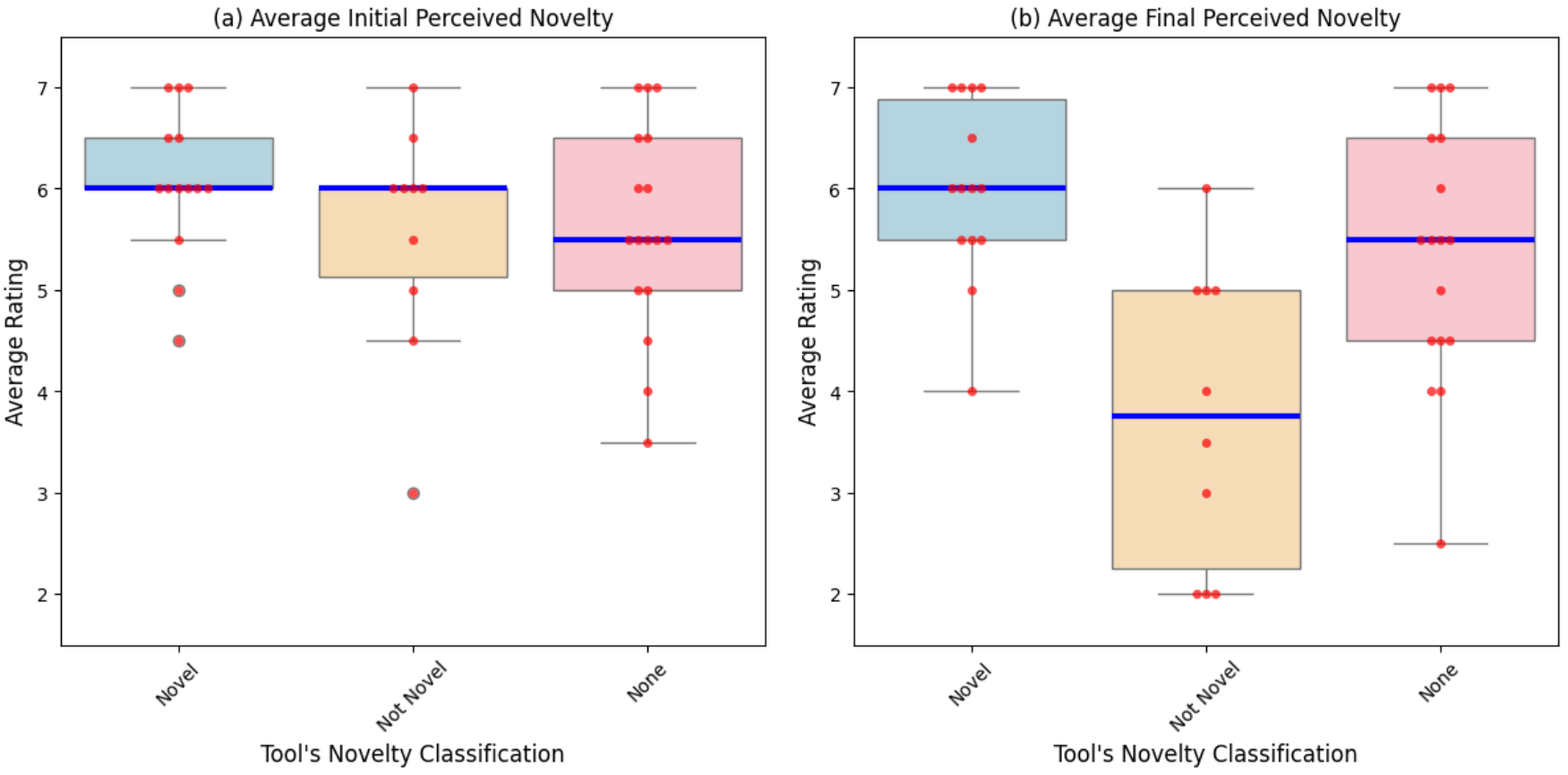}
  \vspace{-15mm}
  \caption{Participants' average perceived idea novelty before (a) and after (b) utilizing their assigned tool for idea novelty evaluation.}
  \label{fig:evalBoxplots}
  \vspace{-4mm}
\end{figure}

Among the 17 participants who completed the novelty evaluation task with the intended setup,\footnote{One participant did not receive the full allotted time; four experienced a different version of the novelty checker due to an API issue.} \sysname's novelty checker did not significantly improve overall confidence in novelty assessments (sign test, S=5.00, p=n.s.). However, a between-subjects comparison revealed that participants changed their novelty assessments most when \sysname\ classified an idea as ``not novel'' (Figure~\ref{fig:evalBoxplots}), suggesting the checker is most useful for filtering unoriginal ideas. This reflects an asymmetry in verifiability: a ``not novel'' classification can be checked against the retrieved papers, whereas a ``novel'' classification is harder to confirm. P17-HCI-treatment captured this asymmetry: 
\aptLtoX{\begin{mdframed1}``\textit{Seeing a list of related work is very helpful for giving you the context. It was very convincing in the case of telling me that an idea was not novel.... 
When it provides [a novel classification], it’s less convincing but is helpful.}''\end{mdframed1}}{\begin{mdframed}[
  backgroundcolor=gray!10,
  leftline=true,
  rightline=false,
  topline=false,
  bottomline=false,
  linecolor=black!40,
  linewidth=2pt,
  innerleftmargin=6pt,
  innerrightmargin=4pt,
  innertopmargin=3pt,
  innerbottommargin=3pt, 
  skipabove=10pt,
  skipbelow=10pt
]
``\textit{Seeing a list of related work is very helpful for giving you the context. It was very convincing in the case of telling me that an idea was not novel.... 
When it provides [a novel classification], it’s less convincing but is helpful.}''
\end{mdframed}}
Three participants went further, noting that the retrieved papers were more useful than the classification or reasoning alone. P8-NLP-treatment explained why: 
\aptLtoX{\begin{mdframed1}``\textit{I didn't really pay much attention to these reasons because reasons can be kind of made up to explain why their generation is novel. So, I kind of relied more on the references that it retrieved.}''\end{mdframed1}}{\begin{mdframed}[
  backgroundcolor=gray!10,
  leftline=true,
  rightline=false,
  topline=false,
  bottomline=false,
  linecolor=black!40,
  linewidth=2pt,
  innerleftmargin=6pt,
  innerrightmargin=4pt,
  innertopmargin=3pt,
  innerbottommargin=3pt, 
  skipabove=10pt,
  skipbelow=10pt
]
``\textit{I didn't really pay much attention to these reasons because reasons can be kind of made up to explain why their generation is novel. So, I kind of relied more on the references that it retrieved.}''
\end{mdframed}}

\subsection{Novelty Checker Ablations}
\label{sec:retrieval_ablation}

We run two ablations to isolate the contributions of the novelty checker's main components: a \textit{retrieval ablation} (Section~\ref{sec:retrieval_ablation_results}) changing only the retrieval pipeline with the classifier held constant, and a \textit{classifier ablation} (Section~\ref{sec:classifier_ablation}) holding retrieval constant and changing only the classification prompt and in-context examples. Together they test whether retrieval and classification each independently benefit from the faceted representation.

\subsubsection{Retrieval Ablation}
\label{sec:retrieval_ablation_results}

\noindent\textbf{Setup.} We evaluate whether the retrieval pipeline surfaces the papers needed to identify ``not novel'' ideas. We focus on ``not novel'' ideas because each has known overlapping papers, so we can check whether each retrieval variant finds them. We exclude ``novel'' ideas because we have no ground truth for them. Our test set contains 58 ideas with known overlapping prior work: 13 ``not novel'' instances from the held-out set (Section~\ref{sec:ideanovelty_module}) plus 45 published NLP papers. The classifier prompt and in-context examples are held constant; only the retrieval pipeline varies. We use \textsf{o3-mini} for novelty classification and \textsf{gpt-4o} for re-ranking, comparing five configurations: (i) \textbf{Complete System} — keyword + snippet retrieval, embedding filtering, facet-based RankGPT; (ii) \textbf{Relevance RankGPT} — facet-based re-ranker replaced with general-relevance re-ranking~\cite{sun2023chatgpt}; (iii) \textbf{Embedding Filtering} — no LLM re-ranker; (iv) \textbf{Snippet Retrieval} — top-$k$ from snippet search only; (v) \textbf{Keyword Retrieval} — top-$k$ from keyword search only.

\begin{table}[ht]
\centering
\small
\caption{Retrieval ablation.}

\label{tab:class_ablation}
\begin{tabular}{lc}
\toprule
\textbf{Method} & \textbf{Accuracy} \\
\midrule
Complete System           & \textbf{89.66\%} \\
- Relevance RankGPT       & 13.79\% \\
- Embedding Filtering     & 10.34\% \\
- Snippet Retrieval       & 8.62\%  \\
- Keyword Retrieval       & 5.17\%  \\
\bottomrule
\end{tabular}

\end{table}

\noindent\textbf{Results.} Table~\ref{tab:class_ablation} shows the complete system substantially outperforms all variants. Keyword- and snippet-only retrieval surface key papers far less reliably, and even alternate re-ranking strategies (embedding-only, or embedding plus general-relevance RankGPT) fail to consistently bring overlapping work into the top 10. Combining facet-based re-ranking with embedding filtering is critical for retrieval that supports novelty assessment.

\subsubsection{Classifier Ablation}
\label{sec:classifier_ablation}

\textbf{Setup.} The retrieval ablation above varies retrieval while holding the classifier constant. To isolate the classifier's contribution, we run a complementary ablation that holds retrieval constant and varies only the classification prompt. Using a held-out test set (26 novel, 17 not novel) from the expert annotation study (Section~\ref{sec:ideanovelty_module}), where annotators reviewed ideas using the top-10 retrieved papers, we fix those top-10 papers and compare four classifier configurations: (i) a zero-shot baseline; (ii) the AI Scientist novelty prompt~\cite{lu2024ai}; (iii) in-context examples drawn from OpenReview idea/review pairs; and (iv) our facet-grounded prompt with expert-labeled in-context examples, with and without explicit reasoning.

\begin{table}[ht]
\centering
\caption{Classifier ablation}
\label{tab:classifier_ablation}
\small
\begin{tabular}{lcc}
\toprule
\textbf{Method} & \textbf{Accuracy} & \textbf{Macro F1} \\
\midrule
Zero-shot                                                          & 0.68 & 0.65 \\
AI Scientist prompt~\cite{lu2024ai}                                & 0.47 & 0.44 \\
OpenReview in-context examples                                     & 0.59 & 0.43 \\
Expert ICL (idea + papers + class)                                 & 0.78 & 0.77 \\
Expert ICL (idea + papers + class + reasoning)                     & \textbf{0.81} & \textbf{0.79} \\
\bottomrule
\end{tabular}

\end{table}

\noindent\textbf{Results.} Table~\ref{tab:classifier_ablation} shows that our facet-grounded classifier substantially outperforms zero-shot, the AI Scientist prompt, and OpenReview-based in-context examples on the same retrieved papers. Two findings are worth highlighting. First, expert-labeled examples with facet-level reasoning outperform OpenReview examples, indicating that \textit{how} novelty is described in the in-context examples matters beyond simply having examples. Second, decomposing ideas and papers into facets outperforms prompts that reason over ideas and papers as wholes, showing that the faceted representation contributes at the classification stage and not only at retrieval.

\medskip

\noindent Together, the retrieval and classifier ablations show that the faceted representation independently benefits both components of the novelty checker. We next investigate whether it also benefits the human-facing interaction.

\section{Related Work}\label{sec:relatedwork}

This work builds upon prior work on facet-based ideation, scientific idea novelty evaluation, and the challenges of human-LLM systems for scientific ideation.

\subsection{Facet-Based Ideation}
\label{sec:haifacetrw}

Combinatorial creativity --- combining existing concepts into new ideas --- accounts for a substantial portion of scientific progress~\cite{boden2004creative,thagard2012cognitive}, and a key approach to supporting it is through combining idea facets such as purpose (the problem) and mechanism (the solution) through analogies across contexts~\cite{holyoak1996mental,hope2017accelerating}.
This framework has facilitated analogy identification across research papers~\cite{chan2018solvent, kang2022augmenting}, product ideas~\cite{hope2017accelerating, hope2022scaling}, biological designs~\cite{kang2024biospark}, and research-paper authors~\cite{portenoy2022bursting}.

Several tools build on structured representations, 
each addressing a different stage and domain. For 
\textit{retrieval}, SOLVENT uses purpose-mechanism 
annotations to help users find analogous 
papers~\cite{chan2018solvent}, with subsequent work 
adding LLM-based facet 
extraction~\cite{kang2022augmenting, portenoy2022bursting}. 
For \textit{exploration}, Luminate extracts response 
dimensions from LLM outputs and lets users navigate 
the design space through structured 
handles~\cite{suh2024luminate}. For 
\textit{recombination}, BIOSPARK supports biological 
analogies in engineering~\cite{kang2024biospark}, 
AnalogiLead supports design 
problems~\cite{srinivasan2024improving}, and 
CreativeConnect supports graphic design through 
keyword recombination~\cite{choi2024creativeconnect}.  Recently, IdeaSynth \cite{pu2024ideasynth} supports refinement of an existing idea via coarse-grained sections (background/method/findings), but does not support early ideation from a few papers, multiple idea directions, recombination across papers, or novelty assessment. Across all these tools, the representation supports a single stage and does not propagate across the pipeline or connect generation to evaluation.

\sysname\ propagates the same faceted 
representation --- purposes, mechanisms, and 
evaluations --- across retrieval, generation, novelty 
evaluation and suggestion. Because every paper and 
idea are described in the same terms, users can 
compare across dozens of directions quickly; 
and a facet selection during exploration directly 
shapes which ideas are generated, which papers are 
retrieved for novelty assessment, and what 
improvements are suggested.

\subsection{Scientific Idea Novelty Evaluation}
\label{sec:noveltyevalrw}

Evaluating whether a generated idea is novel 
relative to existing literature is particularly 
challenging when ideas recombine concepts from 
unfamiliar sub-areas~\cite{dean2006identifying}. 
Fully automated systems like AI 
Scientist~\cite{lu2024ai} retrieve papers via 
keyword similarity and iteratively compare them 
against the idea, but without structured comparison 
of which facets specifically overlap. In human-LLM 
interaction, Acceleron~\cite{nigam2024acceleron} 
uses agent personas to assess and improve a 
proposal's novelty, and 
ReviewFlow~\cite{sun2024reviewflow} provides 
in-situ novelty support for peer reviewers. However, 
both evaluate externally provided ideas and cannot 
connect evaluation back to generation. Other tools 
surface related papers alongside generated 
ideas~\cite{liu2024ai, liu2024personaflow, pu2024ideasynth} 
but do not provide explicit novelty judgments 
grounded in those papers. Moreover, none of 
these systems have been systematically evaluated for novelty assessment --- existing evaluations either focus on the broader tool rather than the novelty component specifically, or rely on small-scale 
qualitative demonstrations rather than controlled comparisons against expert judgments.  

\sysname\ closes 
the loop: when an idea is classified as ``not 
novel,'' the system provides a rationale referencing 
specific retrieved papers and suggests alternatives 
by replacing individual facets, feeding evaluation 
back into generation. We 
evaluate this component both through automated 
ablation against expert labels and through a user 
study examining how researchers interpret and act on 
novelty assessments in practice.

\subsection{Human-LLM Scientific Ideation Systems}
\label{sec:haifacetscientificrw}

Several works have explored fully automating 
scientific ideation~\cite{lahav2022search, 
baek2024researchagent, wang2023scimon, gu2024llms}, 
but automated methods remain insufficient for 
formulating novel, impactful research 
ideas~\cite{hope2023computational, jansen2025codescientist}. 
This has motivated human-LLM 
tools~\cite{yang2016creative, gottweis2025towards}: 
CoQuest supports divergent exploration through 
plain-text feedback~\cite{liu2024ai}, PersonaFlow 
supports convergent development through persona-driven 
feedback~\cite{liu2024personaflow}, and Perspectra~\cite{liu2025perspectra} lets users steer 
multi-agent deliberation with domain-expert personas, giving users control over \textit{which} agents contribute. In all cases, the human 
interacts through free-text prompts or by selecting 
among LLM outputs --- there is no structured 
representation through which user choices propagate 
across modules. As a result, intent must be 
re-articulated at each stage, and neither user nor 
system can trace how a specific input influenced a 
specific output. This is concerning given evidence that unconstrained 
LLM generation reduces idea 
diversity~\cite{doshi2024generative, meincke2025chatgpt}.  

In \sysname, users can steer generation through facet-level actions --- selecting a purpose, swapping a mechanism, adopting a novelty suggestion --- or let 
the system select facets and step in only when they want to redirect. In either case, interactions propagate as structured signals across all modules, rather than free-text prompts that the LLM must reinterpret at each stage, giving the system precise signals about 
what to change and giving the user traceability over how their choices shaped the output.

\section{Conclusion and Discussion}\label{sec:discussion}

We presented \sysname, a human-LLM compound system that  maintains a single faceted representation across  retrieval, generation, and novelty evaluation for  scientific ideation. 
At each stage, the system does the heavy lifting while the user steers --- selecting facets, choosing ideas, interpreting novelty assessments --- or lets the system make selections and step in when they want to redirect. Throughout, judgment remains with the user: the system proposes, the user decides. Faceted representations are known as fundamental in ideation processes; \sysname\ is the first human-LLM system for facet-based scientific ideation and novelty assessment. 

Our results show that this representation benefits both (a) \textit{users}, who 
reported discovering new concepts through the system's facets as well as improved idea exploration and expressiveness; and (b) the \textit{system}, in which facet-based re-ranking and classification substantially outperformed traditional novelty assessment baselines.

\paragraph{\textbf{Faceted interaction extends users' conceptual reach beyond their starting papers.}} Aligned with our goal of supporting divergent ideation~\cite{runco2010divergent,cropley2006praise}, our within-subjects study found participants experienced significantly more creativity support with \sysname\ than the baseline, particularly in exploration --- the factor they rated most important. The most validating finding is where new concepts came from: with \sysname, participants attributed them to the tool's facets and generated ideas, while with the baseline, they cited the input papers. Through the faceted representation, concepts from distant papers became accessible without requiring users to read those papers in full.

\paragraph{\textbf{The shared faceted representation benefits both users and system modules.}} The faceted representation served a dual function: it was the vocabulary participants used to steer the system and the data structure the system's modules operated on. User actions (selecting a purpose, swapping a mechanism) propagated directly to downstream modules without translation, and system-generated ideas composed from selected facets
were immediately interpretable to users. 
Furthermore, the novelty module performed substantially better with facet-based retrieval and classification. The novelty checker was most convincing when it classified ideas as ``not novel'' --- participants could verify the judgment against retrieved papers --- while ``novel'' classifications were harder to trust. Prior work has explored structured representations for individual stages of creative support~\cite{suh2024luminate,chan2018solvent,pu2024ideasynth}; our results suggest that maintaining a consistent representation across a full compound pipeline amplifies these benefits.

\paragraph{\textbf{Implications and risks.}} \sysname's faceted interaction is designed to \emph{support} scientific thinking, not replace it. Free-form LLM generation has been shown to reduce idea diversity~\cite{doshi2024generative,padmakumar2024writing,anderson2024homogenization} as models default to common patterns in their training data. \sysname\ counters this by grounding ideation in retrieved papers from a range of conceptual distances, surfacing cross-area inspirations that users would not have considered on their own. But the same system can also be misused: researchers may offload thinking, over-rely on LLM-generated facets and novelty assessments, or accept outputs without verification. Our facet-level transparency and human-in-the-loop framing are deliberate mitigations, but they do not eliminate these risks.

\paragraph{\textbf{Future directions.}} Several avenues remain open. Participants sometimes avoided distant facets, which future work could support through in-situ question-answering about the unfamiliar facets, or by surfacing the analogy that help scientists engage with these concepts. Some participants also valued paper-level interaction alongside facets, suggesting that tools offering multiple levels of granularity may lead to richer interactions. The novelty checker's coverage could also be improved with adaptive or broader retrieval, since 10 retrieved papers sometimes missed important works. \sysname's modules currently operate in a fixed pipeline directed by the user; a natural extension is to give modules more autonomy, for example having the novelty checker proactively suggest facet replacements during generation.

\bibliographystyle{ACM-Reference-Format}
\bibliography{refs}

\appendix

\section*{Appendix Overview}
\label{app:overview}

\noindent This appendix provides supplementary material organized into the following sections:

\vspace{3mm}

\aptLtoX{{\textbf{A}}\,\,\,\,{User Study: Design,  Procedure, Participants}

{}

{\textbf{B}}\,\,\,\,{User Study: Extended Results}
{}

{\textbf{C}}\,\,\,\,{Annotation Study for System Modules}
{}

{\textbf{D}}\,\,\,\,{Novelty Checker Additional Ablation Results}
{}


{\textbf{E}}\,\,\,\,{Sample Generated Ideas}
{}

{\textbf{F}}\,\,\,\,{Implementation Details}
{}

{\textbf{G}}\,\,\,\,{System Figures: UI and individual components}
{}

{\textbf{G}}\,\,\,\,{LLM Prompts}
{}}{\appentry{\textbf{A}}{User Study: Design,  Procedure, Participants}
{\pageref{app:studydesign}}
{}
\appentry{\textbf{B}}{User Study: Extended Results}
{\pageref{app:extended_results}}
{}

\appentry{\textbf{C}}{Annotation Study for System Modules}
{\pageref{app:module_annotation}}
{}

\appentry{\textbf{D}}{Novelty Checker Additional Ablation Results}
{\pageref{app:nc_retreival_overlap}}
{}


\appentry{\textbf{E}}{Sample Generated Ideas}
{\pageref{sec:longIdeas}}
{}

\appentry{\textbf{F}}{Implementation Details}
{\pageref{app:implementation}}
{}

\appentry{\textbf{G}}{System Figures: UI and individual components}
{\pageref{fig:treatmentevaluation}}
{}

\appentry{\textbf{G}}{LLM Prompts}
{\pageref{app:prompts}}
{}}

\section{User Study: Design and Procedure}
\label{app:studydesign}

\subsection{Participants}
\label{app:participants}
We recruited 22 computer-science researchers (W: 7, M: 15) through institutional mailing lists and academic social networks. We compensated them with a \$60 Amazon gift card. Twelve participated as human-computer interaction (HCI) researchers and 10 as natural-language-processing (NLP) researchers. Most were PhD students (PhD student: 16, master's student: 5, industry researcher: 1). Generally, the participants interacted with LLMs often (a few times per... day: 12, week: 7, month: 1, few months or longer: 2).

\subsection{Study Design}
\label{app:studydesign_details}
We conducted a within-subjects study, in which each participant completed tasks for the treatment and baseline conditions in randomized order. The ideation topics for the treatment and baseline conditions were also randomized. Overall, participants had no difference in their familiarity ratings (7-point, Likert-type) for the assigned treatment topic and assigned baseline topic (M=0.00, Q1=-1.00, Q3=1.00). There were two preset topics for HCI researchers (human-AI collaboration in art, AI tools for education) and two for NLP researchers (dealing with LLM hallucinations, LLM explainability). For each topic, there were three associated input papers to use as a starting point. The input papers for each topic are listed in the following Table.

\begin{table}[H]
\centering
\small
\label{tab:input-papers}
\vspace{-2mm}
\begin{tabularx}{\columnwidth}{@{} p{2.7cm} X @{}}
\toprule
\textbf{Topic} & \textbf{Input Papers} \\
\midrule
Human-AI Collaboration in Art &
LumiMood \cite{oh2024lumimood}; 
Prompting for Discovery \cite{almeda2024prompting}; 
Algorithmic Ways of Seeing \cite{meyer2024algorithmic} \\

AI Tools for Education &
VIVID \cite{choi2024vivid};

Scientific and Fantastical \cite{cheng2024scientific}; 
Putting Things into Context \cite{leong2024putting} \\

\hline

Dealing with LLM Hallucinations &
Deductive Closure Training \cite{akyurek2024deductive}; 
Self-Alignment for Factuality \cite{zhang2024self}; 
HALoGEN \cite{ravichander2025halogen} \\

LLM Explainability &
LLMFactor \cite{wang2024llmfactor}; 
TextGenSHAP \cite{enouen2023textgenshap}; 
Digital Socrates \cite{gu2023digital} \\
\bottomrule
\end{tabularx}
\end{table}

\subsection{Baseline Tool}
\label{app:baseline}
\begin{figure*}[p]
   \centering
   \includegraphics[width=\textwidth]{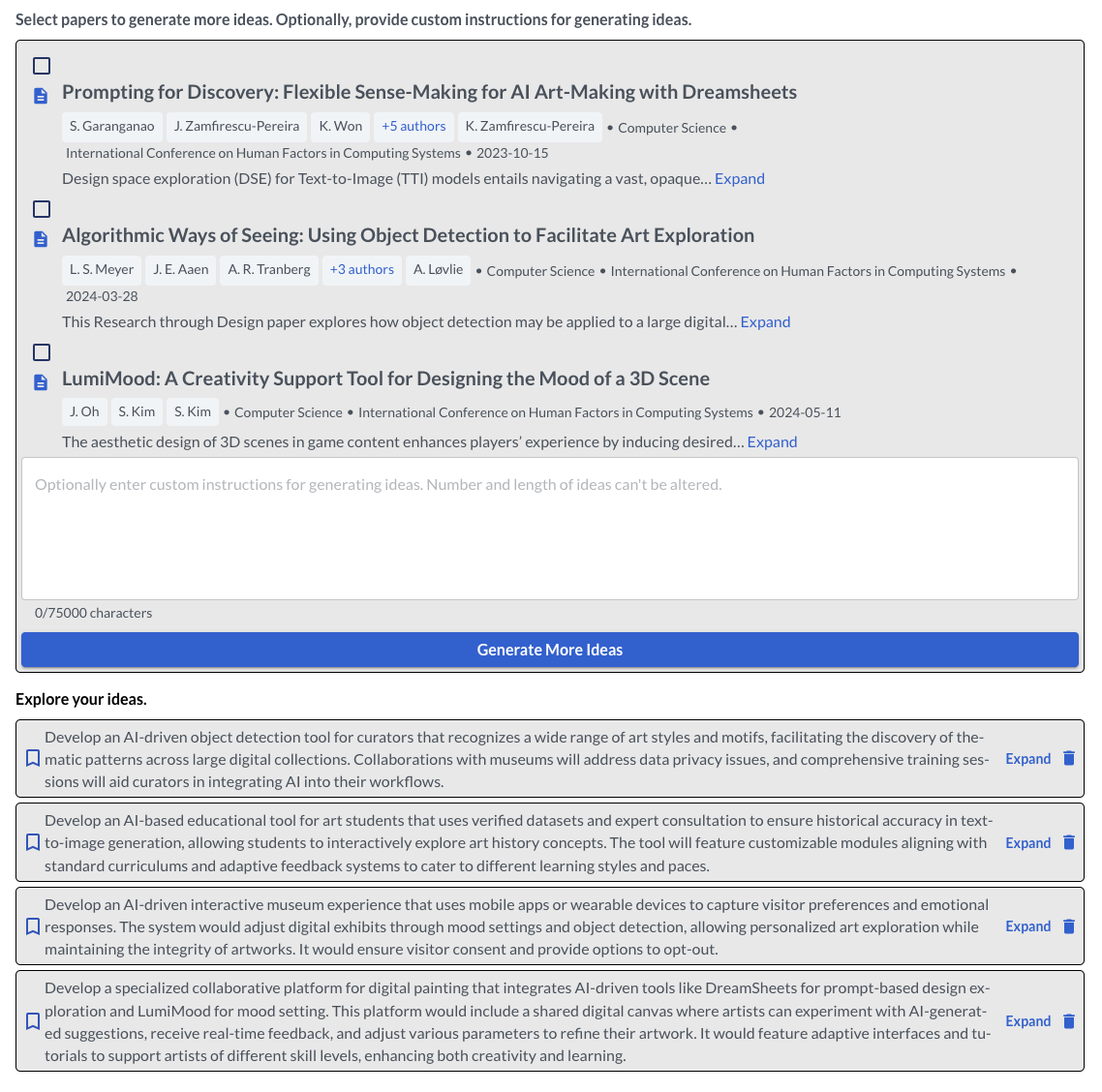}
  \centering
  \caption{The cold start of the baseline UI for the user study's idea-generation task. The ideation topic here is human-AI collaboration in art.}
  \Description{A screenshot. At the top, there is text saying "Select papers to generate more ideas. Optionally, provide custom instructions for generating ideas. Below, there is a gray box with three checkable (but currently unchecked) paper titles. Under each paper title are the paper's authors, field, venue, and first line of the abstract (with an "expand" button to expand it. At the bottom of the gray box, there is a text field with the placeholder text "Optionally enter custom instructions for generating ideas. Number and length of ideas can't be altered." The text field shows how much of the character limit has been taken for the text field ("0/75000 characters"). Below, there is a button labeled "Generate More Ideas." Below, there is text saying "Explore your ideas." Below that, there are four gray boxes, each with text about 75 words long, an unpressed bookmark icon button on the left, an "Expand" button on the right, and a trash icon button to the right.}
  \label{fig:baseline}
\end{figure*}

The baseline tool (Figure~\ref{fig:baseline}) represents \textit{paper-level interaction}: participants could select any combination of the three input papers as input to the LLM \textsf{gpt-4o-2024-08-06}, the same LLM used for most of \sysname's functionality. If they did not select any papers, all three were provided to the LLM. Participants could also provide custom instructions to the LLM, with a character limit of 75,000 (compared to 25,000 in the treatment) to account for the treatment tool's longer set idea-generation prompt. The baseline's idea-generation prompt was a simplified version of the one in \sysname: it did not utilize any facet-based framework or carefully crafted criteria for a good idea. However, like \sysname, it generated six candidate ideas for every two presented to the participant and followed instructions to improve upon the idea. The baseline's ``Idea Novelty Evaluation'' tab was similar to that in the treatment tool except there was no Idea Novelty Checker module output (i.e., no related papers, novelty classification, or classification reason for each idea).

\subsection{Treatment Tool Modifications for Study}
\label{app:treatment_modifications}
We modified \sysname\ to more effectively address our research questions. Our study separates the idea generation task from the idea evaluation task. To keep the study controlled, we disabled some of \sysname's functionalities: on-demand novelty evaluation, manual idea addition, and facet generation when there is no query. The Idea Novelty Checker module was only activated in a separate ``Idea Novelty Evaluation'' tab for the idea evaluation step. There was no support for adjusting the novelty assessment or iterating on the idea's novelty. The tab also provided access to a ChatGPT-like interaction in which participants could prompt the LLM directly to help evaluate their ideas for novelty, as well as a text field for keeping notes on their novelty assessments.

\subsection{Procedure}
\label{app:procedure}
Each within-subjects study session was 105 minutes. The sessions were recorded and transcribed using Google Meet. In each condition, the session coordinator provided the participant with the assigned tool, a document with the titles and abstracts of the input papers for the assigned ideation topic, and a link to Semantic Scholar.\footnote{https://www.semanticscholar.org/} They had access to these three resources throughout the condition. The participant completed two tasks with each tool: an idea-generation task followed by an idea-novelty-evaluation task.

\paragraph{\textbf{Idea-generation task.}} The participant entered their assigned ideation topic and three input papers into the tool. While the tool loaded, the coordinator went over the task instructions and gave the participant a tutorial describing the tool's features. The participant then had up to two minutes to review the three input papers' titles and abstracts. With access to the tool, Semantic Scholar, and the input paper document, the participant subsequently spent 20 minutes generating and saving as many research ideas as possible. To save an idea, the participant had to confirm that the idea was at least somewhat relevant to the ideation topic and somewhat interesting to think about further. They also provided a seven-point Likert-type rating of how different the idea was from ideas they had or encountered before the study; they were told to aim for saving ideas that were at least somewhat significantly different. The coordinator alerted the participant when five minutes remained.

\paragraph{\textbf{Idea rating and survey.}} Once 20 minutes had passed, the participant opened a ``Saved Ideas'' tab to select their two favorite ideas and answer additional 7-point Likert-type questions about their perceived novelty, feasibility, specificity, impact, and imaginativeness of each idea. There were two instances in which a participant had only saved one idea in the 20 minutes allotted. In this case, we asked them to select their next favorite idea in order to proceed with two favorite ideas. Participants also rated their confidence in their novelty assessment. They then completed a survey that included seven-point Likert-type questions about their familiarity with the assigned topic and whether they encountered concepts they had not previously heard about or encountered in the context of the ideation topic. The survey also included the Creativity Support Index (CSI) questionnaire~\cite{cherry2014quantifying}. In the survey for the second tool, the participant also answered questions for each pair of CSI factors to determine which factors they considered most important, as is standard for the CSI. The coordinator then spent up to around five minutes engaging the participant in a semi-structured interview about their idea generation experience.

\paragraph{\textbf{Idea-novelty-evaluation task.}} The participant opened the ``Idea Novelty Evaluation'' tab, and the coordinator provided an overview of this portion of the tool. The participant spent five minutes evaluating their two favorite ideas for novelty. For each idea, they provided a final seven-point Likert-type rating of perceived novelty and confidence in their novelty assessment. The coordinator then conducted a brief semi-structured interview about the participant's idea evaluation experience.

\section{User Study: Extended Results}
\label{app:extended_results}

\begin{figure*}[tb]
  \includegraphics[width=\textwidth]{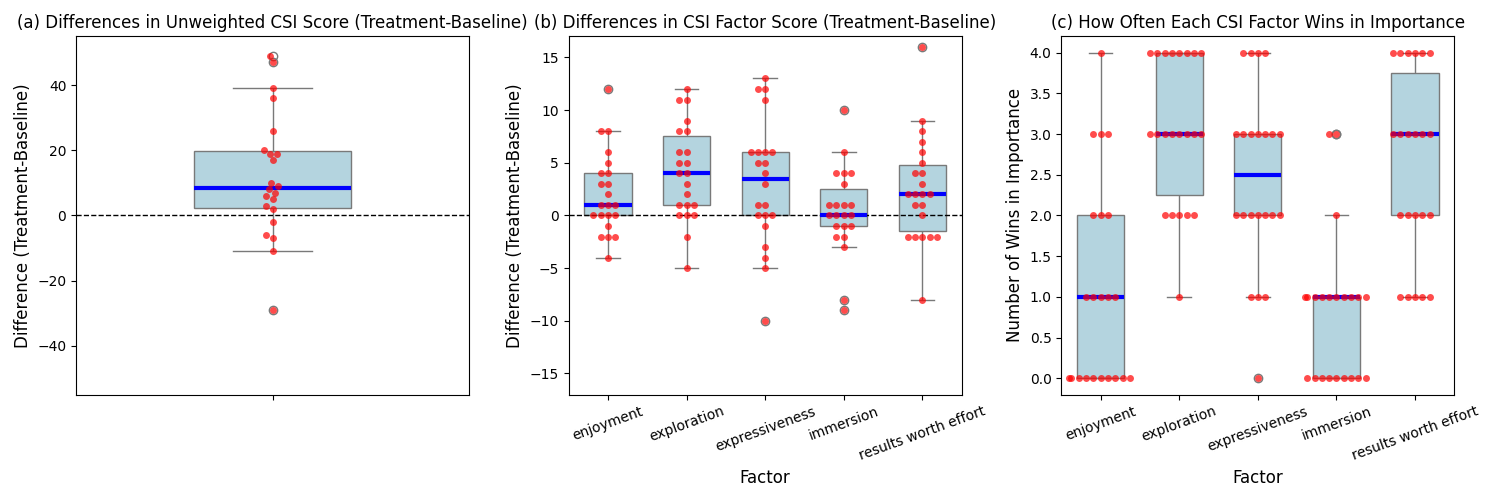}
  \centering
  \caption{(a) The difference between participants' unweighted CSI scores for \sysname\ versus the baseline tool. Participants experienced significantly more creativity support with \sysname. (b) For each CSI factor, the difference between participants' ratings for \sysname\ versus the baseline tool. (c) How many times each CSI factor wins against other factors in terms of what is most important to participants while generating ideas.}
  \Description{(a) A boxplot titled "Differences in Unweighted CSI Score (Treatment-Baseline)." The y-axis is labeled "Difference (Treatment-Baseline)" and has ticks from -40 to 40. The box plot's first quartile is near but above zero, the median is around 7, and the third quartile is around 20. (b) A set of five boxplots titled "Differences in CSI Factor Score (Treatment-Baseline)." The y-axis is labeled "Difference (Treatment-Baseline) and has ticks -15 to 15. The x-axis is labeled "Factor" and has ticks for the five boxplots: enjoyment, exploration, expressiveness, immersion, and results worth effort. For each of those boxplots respectively: 1) Q1 \~ 0, M \~ 1, Q3 \~ 4; 2) Q1 \~ 1, M \~ 4, Q3 \~ 8; 3) Q1 \~ 0, M \~ 4, Q3 \~ 7; 4) Q1 \~ -1, M \~ 0, Q3 \~ 3; 4) Q1 \~ -2, M \~ 3, Q3 \~ 5. Thus, we see that the exploration factor seemed to get the most improvement from the treatment, followed by expressiveness and enjoyment. Little improvement for results worth effort, and essentially no difference for immersion. (c) A set of five boxplots titled "How Often Each CSI Factor Wins in Importance." The y-axis is labeled "Number of wins in importance" and has ticks for 0.0 to 4.0. The x-axis is labeled "Factor" and has ticks for the five boxplots: enjoyment, exploration, expressiveness, immersion, and results worth effort. For each of those boxplots respectively: 1) Q1 \~ 0, M \~ 1, Q3 \~ 2; 2) Q1 \~ 2.25, M \~ 3, Q3 \~ 4; 3) Q1 \~ 2, M \~ 2.5, Q3 \~ 3; 4) Q1 \~ 0, M \~ 1, Q3 \~ 1; 4) Q1 \~ 2, M \~ 3, Q3 \~ 3.75. Thus, participants seemed to find exploration most important, followed by results worth effort and then expressiveness. Then, enjoyment and finally immersion.}
  \label{fig:all_CSI}
\end{figure*}


\subsection{Additional CSI Factor Results}
\label{app:csi_factors}

The main paper reports the exploration and expressiveness factors in detail. Here we report the remaining CSI factors.

\paragraph{Immersion.}
Overall, participants did not find \sysname\ more helpful than the baseline tool for becoming immersed in the idea-generation task. The interviews and interaction logs provide some reasons why this may be true. \sysname\ presents the user with several features about which to learn, and the cognitive demand of learning these features may have prevented immersion. Four participants commented on the high cognitive load of using \sysname; P2-HCI-treatment commented, 
\aptLtoX{\begin{mdframed1}``\textit{I would say that it took me more mental effort to figure out how the tool [is used] rather than work with ideas.}''\end{mdframed1}}{\begin{mdframed}[
  backgroundcolor=gray!10,
  leftline=true,
  rightline=false,
  topline=false,
  bottomline=false,
  linecolor=black!40,
  linewidth=2pt,
  innerleftmargin=6pt,
  innerrightmargin=4pt,
  innertopmargin=3pt,
  innerbottommargin=3pt
]
``\textit{I would say that it took me more mental effort to figure out how the tool [is used] rather than work with ideas.}''
\end{mdframed}}
 Furthermore, due to more complex prompting, the average latency for generating two ideas in \sysname\ was 22.04 seconds, compared to 15.21 seconds in the baseline tool. Ideas were generated in groups of four, two at a time, so the first two ideas took less time to generate than the last two ideas.

\paragraph{Results-Worth-Effort.}
Participants generally found their results to be more worth the effort while using \sysname\ compared to the baseline tool. However, there was little difference in participants' average ratings of their favorite ideas in terms of perceived feasibility (M=0.00, Q1=-0.50, Q3=0.50), specificity (M=0.00, Q1=-0.88, Q3=0.50), and imaginativeness (M=0.00, Q1=-0.50, Q3=0.38). Meanwhile, the baseline tool performed slightly better with respect to generating impactful ideas (M=-0.25, Q1=-0.50, Q3=0.00). Perhaps participants could more clearly see the potential impact of ideas grounded in a few set input papers, which they reviewed, versus ideas drawing from several papers, most of which were not reviewed by them. P10-HCI-baseline posited, 
\aptLtoX{\begin{mdframed1}``\textit{since now I know the paper, I read them, I kind of understand the vocabulary... it is easier for me to see where these ideas are coming from. So even when the ideas are written somewhat vaguely, I can still... imagine how that would pan out because I read the paper.}''\end{mdframed1}}{\begin{mdframed}[
  backgroundcolor=gray!10,
  leftline=true,
  rightline=false,
  topline=false,
  bottomline=false,
  linecolor=black!40,
  linewidth=2pt,
  innerleftmargin=6pt,
  innerrightmargin=4pt,
  innertopmargin=3pt,
  innerbottommargin=3pt
]
``\textit{since now I know the paper, I read them, I kind of understand the vocabulary... it is easier for me to see where these ideas are coming from. So even when the ideas are written somewhat vaguely, I can still... imagine how that would pan out because I read the paper.}''
\end{mdframed}}

\paragraph{Enjoyment.}
Most participants benefited slightly from \sysname\ in terms of enjoyment, but the difference from the baseline was small (Figure~\ref{fig:all_CSI}b).

\subsection{Evaluation Facet Utility}
\label{app:eval_facet}
Participants tended not to find the evaluation facets as helpful as the purpose and mechanism facets. Four participants commented that they found the evaluation facet unimportant. P13-HCI-treatment elaborated, 
\aptLtoX{\begin{mdframed1}``\textit{For evaluation, I really don't think it's necessary for me because once you have the problem, you have the solution. Automatically you know how to evaluate it, like what study you need, what kind of experiment you want to have, and what variables you are measuring.}''\end{mdframed1}}{\begin{mdframed}[
  backgroundcolor=gray!10,
  leftline=true,
  rightline=false,
  topline=false,
  bottomline=false,
  linecolor=black!40,
  linewidth=2pt,
  innerleftmargin=6pt,
  innerrightmargin=4pt,
  innertopmargin=3pt,
  innerbottommargin=3pt
]
``\textit{For evaluation, I really don't think it's necessary for me because once you have the problem, you have the solution. Automatically you know how to evaluate it, like what study you need, what kind of experiment you want to have, and what variables you are measuring.}''
\end{mdframed}}
Future work may investigate whether the evaluation facet is useful for mixed-initiative, facet-based generation of research ideas.


\subsection{Paper-Level Interaction Preferences}
\label{app:paper_level}
Five participants appreciated the baseline's paper-level interactions in addition to or more than \sysname's facet-level interactions. Three of these participants liked the paper-level interaction, as it felt more directly connected to the literature. P22-NLP-baseline explained, 

\aptLtoX{\begin{mdframed1}``\textit{I think papers for me were more natural than facets.... I think to me it's more like a map of literature, so I could see it more with papers.}''\end{mdframed1}}{\begin{mdframed}[
  backgroundcolor=gray!10,
  leftline=true,
  rightline=false,
  topline=false,
  bottomline=false,
  linecolor=black!40,
  linewidth=2pt,
  innerleftmargin=6pt,
  innerrightmargin=4pt,
  innertopmargin=3pt,
  innerbottommargin=3pt
]
``\textit{I think papers for me were more natural than facets.... I think to me it's more like a map of literature, so I could see it more with papers.}''
\end{mdframed}}

Three participants thought a combination of the two tools would be helpful. Two participants even proposed distinct roles for the two tools: divergent-ideation for \sysname\ and convergent-ideation for the baseline. P7-HCI-treatment shared, 

\aptLtoX{\begin{mdframed1}``\textit{In the [baseline tool], I started from a broader view and then I narrowed it down. Here [in \sysname], I started from a very specific thing and then I tried to add new facets or ideas so that I can expand the idea. So you see the other process is elimination process, here I was trying to expand.}''\end{mdframed1}}{\begin{mdframed}[
  backgroundcolor=gray!10,
  leftline=true,
  rightline=false,
  topline=false,
  bottomline=false,
  linecolor=black!40,
  linewidth=2pt,
  innerleftmargin=6pt,
  innerrightmargin=4pt,
  innertopmargin=3pt,
  innerbottommargin=3pt
]
``\textit{In the [baseline tool], I started from a broader view and then I narrowed it down. Here [in \sysname], I started from a very specific thing and then I tried to add new facets or ideas so that I can expand the idea. So you see the other process is elimination process, here I was trying to expand.}''
\end{mdframed}}

\paragraph{Distant Facet Avoidance.}
Relatedly, participants sometimes avoided distant 
facets in \sysname\ because they were unfamiliar with how to 
apply them --- they did not know the facet's meaning, 
found it ``too far'' from their research area, or 
could not see how the mechanism could achieve the 
purpose. Participants saved ideas with far facets 
much less often than ideas with input and near 
facets. By avoiding distant facets, scientists may 
miss opportunities for ideas that would never have 
occurred to them otherwise.

\subsection{Desire for More Papers in Baseline}
\label{app:more_papers}
After using the baseline tool, four participants said that they wanted a way to input more papers to better express themselves, and five more felt limited by the three input papers. P14-NLP-baseline, for example, 
\aptLtoX{\begin{mdframed1}``\textit{would have liked to add a different paper because it felt like I had exhausted... the creativity in the system to some extent.}''\end{mdframed1}}{\begin{mdframed}[
  backgroundcolor=gray!10,
  leftline=true,
  rightline=false,
  topline=false,
  bottomline=false,
  linecolor=black!40,
  linewidth=2pt,
  innerleftmargin=6pt,
  innerrightmargin=4pt,
  innertopmargin=3pt,
  innerbottommargin=3pt
]
``\textit{would have liked to add a different paper because it felt like I had exhausted... the creativity in the system to some extent.}''
\end{mdframed}}
 While participants could add information from papers to their custom instructions, there was no system feature for adding more papers to the list of input papers. Future work may compare \sysname\ with a modified version of the baseline tool that allows users to add as many papers as they want for recombination.

\subsection{Novelty Checker Usage Patterns}
\label{app:novelty_usage}

Participants used the novelty checker an average of 
6 times per session. The checker was most convincing 
when it classified an idea as ``not novel'' --- 
participants could directly verify the judgment 
against the retrieved papers. ``Novel'' 
classifications were less persuasive, as the absence 
of overlapping work is harder to trust than its 
presence. Two common concerns were that the checker 
classified ideas as ``novel'' too often and that the 
set of most related papers sometimes missed important 
works. Both concerns may have been partly due to the 
system retrieving only 10 papers per idea in the user 
study, a constraint imposed by latency limitations. 
Broader retrieval and richer paper representations 
may improve evaluation quality and user trust.

\section{Annotation Study for Modules}
\label{app:module_annotation}

\subsection{Facet Distance Consistency}
\label{app:distance_consistency}

To assess whether the system's distance labels (near, far, very far) correspond to meaningful differences in conceptual distance, the first two authors independently annotated purpose-mechanism pairs generated by the system for three papers not used in the user study.\footnote{This annotation used an earlier but similar version of the tool compared to the version used in the study.} Facets were grouped into two broad categories --- generally near and generally far --- and each annotator classified whether the system's labels matched their own judgment. Both annotators classified the majority of near purposes, near mechanisms, far purposes, and far mechanisms consistently with the system's labels. 
These results suggest the distance levels capture meaningful variation in conceptual similarity, although we did observe certain degree of subjectivity of judging how far is an analogy. 

\subsection{Novelty Checker: Expert-Labeled Examples}
\label{app:expert_examples}

The Idea Novelty Checker uses expert-labeled examples as in-context demonstrations for its novelty assessment step (Section~\ref{sec:ideanovelty_module}). Each example includes the idea (with key facets bolded), the top-10 most relevant papers from the re-ranking pipeline, and the expert's grounded reasoning. These demonstrations teach the LLM to cite specific retrieved papers when justifying its classification. Figure~\ref{fig:expert-novel} shows a novel example and Figure~\ref{fig:expert-notnovel} shows a not-novel example.

\begin{figure*}[hbp!]
\begin{imageonly}
\begin{mdframed}[
  backgroundcolor=green!3,
  linecolor=green!40!black,
  linewidth=0.5pt,
  roundcorner=2mm,
  innerleftmargin=5pt,
  innerrightmargin=5pt,
  innertopmargin=4pt,
  innerbottommargin=4pt,
  frametitlebackgroundcolor=green!20,
  frametitlerule=false,
  frametitleaboveskip=4pt,
  frametitlebelowskip=4pt,
  frametitlefont=\small\bfseries,
  frametitle={Example 1 --- Classification: Novel}
]

\small
\textbf{Idea:} Develop a \textbf{natural language processing classifier designed to improve scientific paper revisions} by automatically identifying and categorizing reviewer comments that are most likely to lead to substantial and actionable revisions. The system would be trained on a \textbf{manually-labeled dataset analysis} of scientific review comments and the corresponding paper edits, leveraging features such as linguistic cues, sentiment, and comment specificity to predict the likelihood of a comment being acted upon. This classifier could then be used to prioritize reviewer feedback, helping authors focus on the most impactful suggestions first.

\vspace{3mm}
\textbf{Top-10 Retrieved Papers:}
\vspace{-1mm}
\begin{enumerate}[leftmargin=1.5em, itemsep=0pt, parsep=0pt, topsep=2pt]
  \item \footnotesize ARIES: A Corpus of Scientific Paper Edits Made in Response to Peer Reviews
  \item \footnotesize Can Large Language Models Provide Useful Feedback on Research Papers?
  \item \footnotesize A Dataset of Peer Reviews (PeerRead): Collection, Insights and NLP Applications
  \item \footnotesize arXivEdits: Understanding the Human Revision Process in Scientific Writing
  \item \footnotesize Characterizing Text Revisions to Better Support Collaborative Writing
  \item \footnotesize Can We Automate Scientific Reviewing?
  \item \footnotesize DeepReviewer: Collaborative Grammar \& Innovation Neural Network for Automatic Paper Review
  \item \footnotesize Aspect-based Sentiment Analysis of Scientific Reviews
  \item \footnotesize Aspect-based Sentiment Analysis of Online Peer Reviews and Prediction of Paper Acceptance
  \item \footnotesize ReviVal: Towards Automatically Evaluating the Informativeness of Peer Reviews
\end{enumerate}

\vspace{2mm}
\textbf{Expert Reasoning:} The idea is \textbf{novel} because it uniquely focuses on \textit{prioritizing} reviewer comments for actionable revisions, which is not explicitly addressed in ARIES~[1] or other related works like ReviVal~[10].
\end{mdframed}
\end{imageonly}
\caption{Expert-labeled \textbf{novel} example used as an in-context demonstration. The retrieved papers address related aspects of peer review (corpora, automation, sentiment) but none specifically tackle the proposed task of \textit{prioritizing} comments by likelihood of leading to revisions --- the gap the expert identifies.}
\label{fig:expert-novel}
\end{figure*}

\begin{figure*}[htp!]
\begin{imageonly}
\begin{mdframed}[
  backgroundcolor=red!3,
  linecolor=red!40!black,
  linewidth=0.5pt,
  roundcorner=2mm,
  innerleftmargin=5pt,
  innerrightmargin=5pt,
  innertopmargin=4pt,
  innerbottommargin=4pt,
  frametitlebackgroundcolor=red!20,
  frametitlerule=false,
  frametitleaboveskip=4pt,
  frametitlebelowskip=4pt,
  frametitlefont=\small\bfseries,
  frametitle={Example 2 --- Classification: Not Novel}
]

\small
\textbf{Idea:} Develop a \textbf{systematic review-based framework} designed \textbf{to align LLM evaluation with human preferences}, ensuring that evaluation criteria are continuously refined based on comprehensive reviews of user feedback and emerging model behaviors. This framework will utilize \textbf{content analysis of user interactions and feedback} to identify patterns and areas of improvement. The effectiveness of this framework will be assessed through a \textbf{qualitative study} involving iterative cycles of user feedback and criteria refinement.

\vspace{3mm}
\textbf{Top-10 Retrieved Papers:}
\vspace{-1mm}
\begin{enumerate}[leftmargin=1.5em, itemsep=0pt, parsep=0pt, topsep=2pt]
  \item \footnotesize EvalLM: Interactive Evaluation of Large Language Model Prompts on User-Defined Criteria
  \item \footnotesize Humanely: Human Evaluation of LLM Yield, Using a Novel Web-Based Evaluation Tool
  \item \footnotesize Evaluation of Code Generation for Simulating Participant Behavior in ESM by Iterative ICL of an LLM
  \item \footnotesize Human-Centered Evaluation and Auditing of Language Models
  \item \footnotesize Aligning Model Evaluations with Human Preferences: Mitigating Token Count Bias in LM Assessments
  \item \footnotesize Who Validates the Validators? Aligning LLM-Assisted Evaluation of LLM Outputs with Human Preferences
  \item \footnotesize Human-Centered Design Recommendations for LLM-as-a-Judge
  \item \footnotesize CheckEval: Robust Evaluation Framework using Large Language Model via Checklist
  \item \footnotesize Discovering Language Model Behaviors with Model-Written Evaluations
  \item \footnotesize Prometheus 2: An Open Source Language Model Specialized in Evaluating Other Language Models
\end{enumerate}

\vspace{2mm}
\textbf{Expert Reasoning:} The idea is \textbf{not novel} because it closely resembles existing frameworks like EvalLM~[1] and HumanELY~[2], which already align LLM evaluations with human preferences using user-defined criteria and human feedback.
\end{mdframed}
\end{imageonly}
\caption{Expert-labeled \textbf{not-novel} example. Unlike Example~1, the top retrieved papers directly overlap with the proposed idea's core contribution --- aligning LLM evaluation with human preferences through iterative feedback --- making the overlap straightforward to identify.}
\label{fig:expert-notnovel}
\end{figure*}

\section{Novelty Checker: Additional Ablation Results}
\label{app:nc_retreival_overlap}

The retrieval ablation in Section~\ref{sec:retrieval_ablation_results} shows that ablated variants retrieve fewer of the papers needed for the classifier to identify ``not novel'' ideas. We further compared the top-10 papers retrieved by each variant against the complete system (Table~\ref{tab:doc_ablation}). Approximately 20--30\% of the papers differ when using either embedding-based filtering or general-relevance RankGPT in place of facet-based re-ranking. In contrast, snippet-only and keyword-only retrieval show minimal overlap with the complete system's top-10, highlighting the importance of the re-ranker stage.

\begin{table}[ht]
\caption{Top-10 papers retrieved by each ablated variant vs. the complete system. \emph{Overlap}: average number of papers shared with the complete system's top-10. \emph{Rank Shift}: average absolute change in rank position among overlapping papers.}
\label{tab:doc_ablation}
\small
\centering
\begin{tabular}{lcc}
\toprule
\textbf{Method} & \textbf{Overlap} ($\uparrow$) & \textbf{Rank Shift} ($\downarrow$) \\
\midrule
Relevance RankGPT     & 7.97  & 0.67  \\
Embedding Filtering   & 7.93  & 0.84  \\
Snippet Retrieval     & 2.88  & 1.85  \\
Keyword Retrieval     & 1.17  & 1.39  \\
\bottomrule
\end{tabular}
\end{table}

\section{Sample Generated Ideas}
\label{sec:longIdeas}

Table~\ref{tab:idea-generation-examples-expanded} presents six sample ideas that participants saved as favorites during the user study---four from \sysname\ and two from the baseline.
For each treatment idea, the table shows which generation method was triggered by the participant's facet selections (Section~\ref{sec:ideagen}), the selected facets with their conceptual distance from the input papers, and any custom instructions the participant provided.
Baseline ideas show only custom instructions, since the baseline uses paper-level interaction rather than facets.
Two patterns are visible: (1)~treatment ideas draw on facets from papers at varying distances, with participants often selecting the purpose themselves while letting the mechanism come from a retrieved paper; and (2)~baseline ideas rely more heavily on custom instructions to steer generation.

\vspace{2mm}

\onecolumn

\begin{table}[h!]
\centering
\small
\renewcommand{\arraystretch}{1.3}
\caption{\textbf{Sample favorite ideas from the user study.} Treatment ideas show the generation method, selected facets (with distance and selection source), and custom instructions. Baseline ideas show only the custom instructions provided. Facets are color-coded: \colorbox{purposegreen}{\strut\,Purpose\,}, \colorbox{mechyellow}{\strut\,Mechanism\,}, and \colorbox{evalpurple}{\strut\,\color{black}Evaluation\,}. ``Selected'' means the participant chose this facet; ``unselected'' means the system chose it automatically.}
\label{tab:idea-generation-examples-expanded}
\vspace{2mm}

\begin{tabular}{@{} p{2cm} p{14cm} @{}}
\toprule

\multicolumn{2}{@{}l}{\cellcolor[HTML]{F7F7F7}\textbf{Idea 1} \hfill \textit{Treatment --- Initial method} \hfill Topic: LLM Explainability} \\
\midrule

\textbf{Facets} &
  \colorbox{purposegreen}{\strut\,to enhance medical information retrieval\,} {\scriptsize(near)} \hspace{6pt}
  \colorbox{mechyellow}{\strut\,sequential knowledge-guided prompting\,} {\scriptsize(input)} \newline
  \colorbox{evalpurple}{\strut\,\color{black}medical QA benchmarks\,} {\scriptsize(near, unselected)} \\[2pt]

\textbf{Custom} & \textit{None} \\[2pt]

\textbf{Idea} & The proposed system focuses on developing a robust medical information retrieval platform using sequential knowledge-guided prompting to simulate diagnostic pathways. By integrating electronic health records (EHRs) with established clinical guidelines, the system leverages prompts to dynamically identify and follow potential diagnostic routes, thereby enhancing the precision and relevance of information retrieval. The process begins by extracting key medical factors from EHRs and matching these with clinical pathway prompts to simulate diagnostic reasoning. This approach allows the system to adjust retrieval strategies based on real-time patient data, ensuring contextually relevant information is provided to healthcare professionals. The system's effectiveness will be rigorously evaluated using medical question-answering benchmarks, which will assess the accuracy and pertinence of information retrieval against standard medical queries. \\

\midrule

\multicolumn{2}{@{}l}{\cellcolor[HTML]{F7F7F7}\textbf{Idea 2} \hfill \textit{Treatment --- P-or-M method} \hfill Topic: Human-AI Collaboration in Art} \\
\midrule

\textbf{Facets} &
  \colorbox{purposegreen}{\strut\,to enhance designer-AI collaboration\,} {\scriptsize(very near, selected)} \hspace{6pt}
  \colorbox{mechyellow}{\strut\,residual learning framework\,} {\scriptsize(far)} \newline
  \colorbox{evalpurple}{\strut\,\color{black}empirical performance analysis\,} {\scriptsize(far, unselected)} \\[2pt]

\textbf{Custom} & \textit{I want to know if designers want to work with AI} \\[2pt]

\textbf{Idea} & In this novel approach, we apply a residual learning framework to AI design tools, treating designer feedback as residuals to refine specific design elements such as color schemes and layout structures. This framework allows designers to provide targeted feedback easily, which the AI tool uses to iteratively improve its suggestions. The process starts with the AI generating an initial design, which the designer can annotate with feedback. This feedback is then incorporated as residuals, prompting the AI to make adjustments that align more closely with the designer's vision. The feedback loop is engineered to be highly intuitive, integrating seamlessly into the designer's workflow, thus fostering a more collaborative and efficient design process. An empirical performance analysis will evaluate the effectiveness of this approach, measuring improvements in design quality and collaboration efficiency. \\

\midrule

\multicolumn{2}{@{}l}{\cellcolor[HTML]{F7F7F7}\textbf{Idea 3} \hfill \textit{Treatment --- P-and-M method} \hfill Topic: LLM Explainability} \\
\midrule

\textbf{Facets} &
  \colorbox{purposegreen}{\strut\,to explain text generation outputs\,} {\scriptsize(input)} \hspace{6pt}
  \colorbox{mechyellow}{\strut\,explanation logic verifier\,} {\scriptsize(own input)} \newline
  \colorbox{evalpurple}{\strut\,\color{black}quantitative and qualitative analysis\,} {\scriptsize(input, selected)} \\[2pt]

\textbf{Custom} & \textit{Use different verification approaches from NLI, math, program to ensure the explanation follows a logical consistency} \\[2pt]

\textbf{Idea} & Create an innovative platform that enhances the validation of LLM-generated mathematical explanations by integrating a mathematical logic verifier with visual proof representations. The platform will employ the verifier to ensure explanations adhere to logical principles, assessing each step for accuracy. Simultaneously, it will generate graph-based visual proofs, providing intuitive visual cues for each stage of the reasoning process. These representations will guide users through complex explanations, making abstract concepts more accessible and engaging. The system will also feature interactive elements that allow users to explore different parts of the proofs, deepening their understanding. This approach will be evaluated through comprehensive quantitative and qualitative analysis, assessing improvements in explanation accuracy, user comprehension, and satisfaction. \\

\midrule

\multicolumn{2}{@{}l}{\cellcolor[HTML]{F7F7F7}\textbf{Idea 4} \hfill \textit{Treatment --- P-and-M method} \hfill Topic: AI Tools for Education} \\
\midrule

\textbf{Facets} &
  \colorbox{purposegreen}{\strut\,to personalize vocabulary learning\,} {\scriptsize(very near)} \hspace{6pt}
  \colorbox{mechyellow}{\strut\,AI-generated virtual speakers\,} {\scriptsize(very near)} \newline
  \colorbox{evalpurple}{\strut\,\color{black}participant feedback analysis\,} {\scriptsize(very near, unselected)} \\[2pt]

\textbf{Custom} & \textit{Make the idea more focused and specific} \\[2pt]

\textbf{Idea} & This innovative platform leverages AI-generated virtual speakers to create personalized vocabulary learning experiences. Users upload personal images, and the AI analyzes these images to identify objects, settings, and potential vocabulary links. The virtual speakers then generate dialogue scenarios around these elements, allowing learners to engage with new vocabulary in a personally relevant context. For example, an image of a beach vacation might lead to dialogues about seaside activities, weather, and local culture, enriching the vocabulary acquisition process. This personalized approach not only aids retention by linking words to familiar contexts but also enhances engagement through interactive AI-driven dialogues. Participant feedback analysis will assess the platform's effectiveness in improving vocabulary acquisition and learner satisfaction. \\

\bottomrule

\end{tabular}
\end{table}

\clearpage

\begin{table}[h!]
\centering
\small
\renewcommand{\arraystretch}{1.3}
\captionsetup{labelformat=empty}
\caption{Table~\ref{tab:idea-generation-examples-expanded} (continued). Baseline ideas.}
\vspace{2mm}
\begin{tabular}{@{} p{2cm} p{14cm} @{}}
\toprule

\multicolumn{2}{@{}l}{\cellcolor[HTML]{F7F7F7}\textbf{Idea 5} \hfill \textit{Baseline} \hfill Topic: Dealing with LLM Hallucinations} \\
\midrule

\textbf{Facets} & \textit{N/A --- baseline uses paper-level interaction} \\[2pt]

\textbf{Custom} & \textit{Can the reward model in Self-Knowledge Tuning be changed to incorporate more fine-grained feedback (e.g., from HALoGEN-like metrics)? [all 3 papers given to LLM]} \\[2pt]

\textbf{Idea} & This research focuses on developing an advanced reward model for Self-Knowledge Tuning by incorporating specific metrics from the HALoGEN framework. The core of this model lies in leveraging atomic unit verification to dissect LLM outputs into fundamental components, evaluating each against reliable knowledge bases. By integrating HALoGEN's error classifications, the model can provide targeted feedback that distinguishes between different types of hallucinations, such as those arising from incorrect recollections, faulty training data, or fabrications. This nuanced feedback is expected to enhance the calibration of LLMs, reducing the frequency of factual inaccuracies. The model's efficacy will be tested across diverse domains, including programming and scientific attribution, to ensure broad applicability. A comprehensive validation framework will be established, involving both automated and human evaluations, to rigorously measure improvements in model accuracy and coherence. \\

\midrule

\multicolumn{2}{@{}l}{\cellcolor[HTML]{F7F7F7}\textbf{Idea 6} \hfill \textit{Baseline} \hfill Topic: AI Tools for Education} \\
\midrule

\textbf{Facets} & \textit{N/A --- baseline uses paper-level interaction} \\[2pt]

\textbf{Custom} & \textit{Generate ideas that use non-traditional mediums for education, so avoid using textbooks or other material commonly found in classrooms. [all 3 papers given to LLM]} \\[2pt]

\textbf{Idea} & This AR-enabled mobile application is designed to engage children with their local environment while providing a robust educational platform about biodiversity and ecology. The AI component will analyze user interactions and location data to curate a personalized educational journey. For a child in an urban environment, the app might focus on urban wildlife and plants, providing detailed information, conservation tips, and interactive quizzes tailored to their locality. For children in rural settings, it might explore more diverse ecosystems, encouraging them to document and learn about various species. \\

\bottomrule
\end{tabular}
\end{table}

\twocolumn

\section{Implementation Details}
\label{app:implementation}

The frontend of \sysname\ was developed using React and TypeScript, and the backend with Python. Unless noted otherwise, we used the LLM \textsf{gpt-4o-2024-08-06} and a temperature of zero. In addition, whenever we mention using a paper, we only use its title and abstract. Prompts for all modules are provided in the Appendix. In facet-idea generator the LLM's temperature is set to 0.75 to make the responses more varied. We use \textsf{gpt-4o} for re-ranking in novelty checker and \textsf{o3-mini} for the novelty assessment step given its stronger reasoning capabilities.

\section{System Figures}
\label{app:system_figures}

\begin{table}[H]
\centering
\small
\caption{Guide to system figures (Appendix~\ref{app:system_figures}).}
\label{tab:system-figure-guide}
\begin{tabularx}{\columnwidth}{@{} p{1.8cm} X @{}}
\toprule
\textbf{Figure} & \textbf{Overview} \\
\midrule
User-Interace & \\
\midrule
Fig.~\ref{fig:treatmentevaluation} 
& \sysname\ novelty assessment UI component: idea, facets, related papers, adjustable novelty rating and rationale, and revision suggestions. \\

Fig.~\ref{fig:treatmentideation} 
& \sysname\ facet-driven ideation interface: select/generate facets, add instructions, browse ideas, and access novelty evaluation. \\

Fig.~\ref{fig:baseline} 
& Baseline ideation UI: select papers, add optional instructions, and browse generated ideas (no facet controls or structured novelty feedback). \\
\midrule
\sysname\ modules & \\
\midrule
Fig.~\ref{fig:facetFinder} 
& Module 1: Analogous Paper Facet Finder \\
Fig.~\ref{fig:noveltyChecker-workflow} 
& Module 3: Idea Novelty Checker \\
\bottomrule
\end{tabularx}
\end{table}

\begin{figure*}[p]
   \centering
   \includegraphics[width=\textwidth]{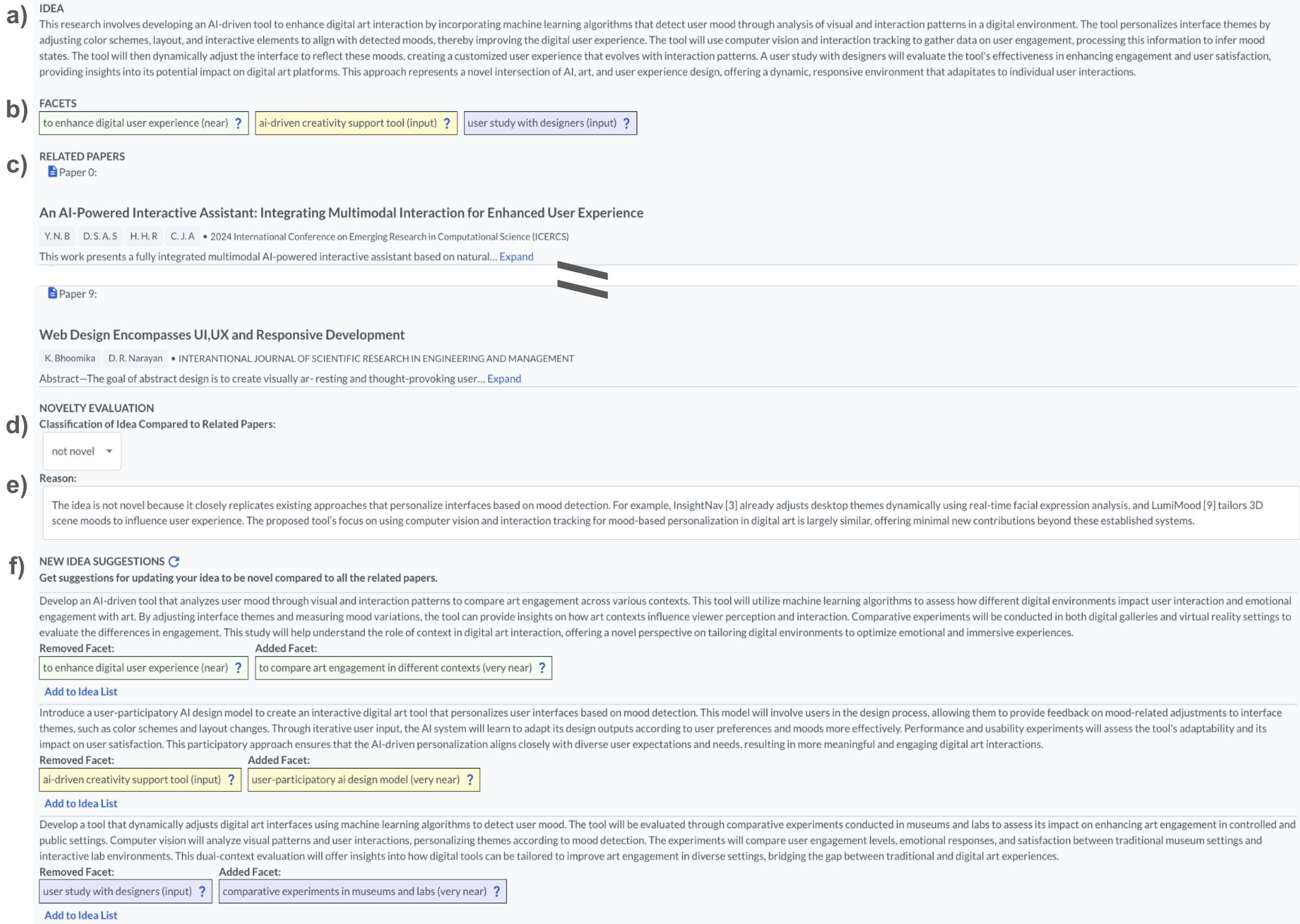}
   \caption{\sysname's novelty assessment modal for one idea, which presents the idea (a) as well as its facets (b), related papers (c), adjustable novelty classification (d), and adjustable classification reason (e). When the idea is classified as ``not novel,'' the system provides a set of three suggestions for more novel ideas (f), each of which replace one of the idea's original facets. The ideation topic here is human-AI collaboration in art.}
   \Description{a) Screenshot with idea text and associated facets in green, yellow, and purple in color-coded boxes below. Below that, the first related paper, then a symbol indicating a jump in screenshot, and then the tenth related paper. Each related paper shown has a title, author, venue, and the first line of the abstract with an "expand" button. At the bottom, novelty evaluation with rating of idea compared to related papers and button to adjust it, which is set to "novel." Reason below is a few sentences with paper citations. Below, new idea suggestions with refresh button saying "Get suggestions for updating your idea to be novel compared to all the related papers. Added and removed facets are indicated for three updated idea suggestions and a button for each says "add to idea list."}
   \label{fig:treatmentevaluation}
 \end{figure*}

\begin{figure*}[p]
   \centering
   \includegraphics[width=\textwidth]{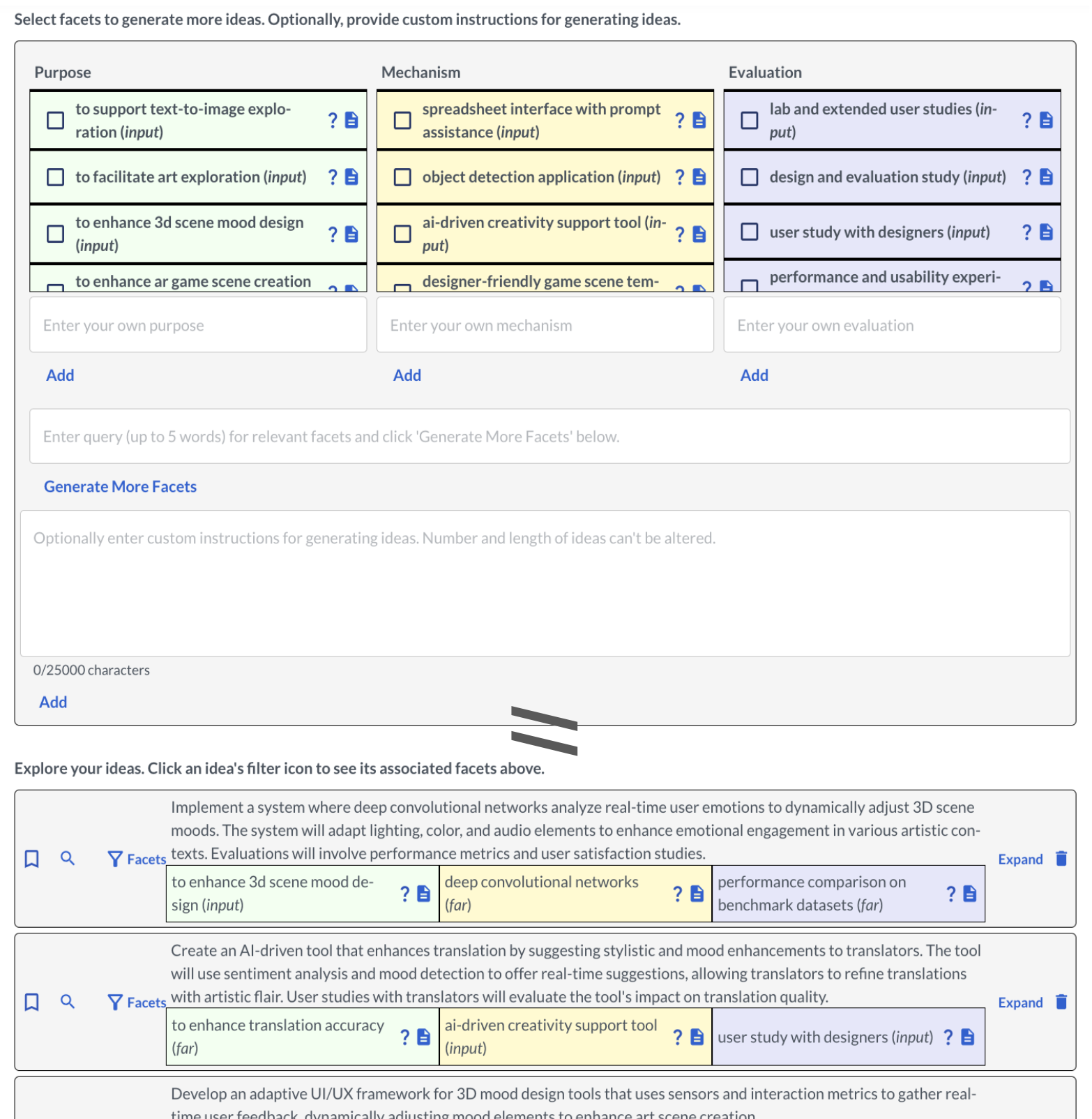}
  \centering
  \caption{\sysname's cold start. Above, the user selects or adds facets to generate ideas. They can also generate more facets to consider, and add custom instructions for the idea generation. Below, the user peruses their ideas and evaluates an idea for novelty by clicking the search icon to its left. The ideation topic here is human-AI collaboration in art.}
  \Description{A screenshot. Top half, from top to bottom, is as follows. "Select facets to generate more ideas. Optionally, provide custom instructions for generating ideas. Below is a gray box with three differently colored columns. A green column labeled "purpose", a yellow column labeled "mechanism", and an evaluation column labeled "purple." A list of facets with checkboxes to the left and question marks and paper icons to the right are included in each column. Below each column are text fields to enter your own facet for the respective column and a button below that says "Add." Underneath is a text field with placeholder text saying "Enter query (up to 5 words) for relevant facets and click 'Generate More Facets' below." Below is a button "Generate More Facets." Below is a text field with the number of characters entered shown (0/25000 characters). The placeholder text for that text field says "Optionally enter custom instructions for generating ideas. Number and length of ideas can't be altered." There is an "Add" button below. Underneath all of this is a button "Generate More Ideas." The bottom half of the screenshot, from top to bottom is as follows. "Explore your ideas. Click an idea's filter icon to see its associated facets above." Below, there is a list of gray boxes with idea text (around 75 words) and tags in green, yellow, and purple. The tags show the idea's facets with their associated question mark and paper buttons. Below is a bookmark icon and search icon to the left of each idea and an "Expand" button and trash icon to the right. The list of ideas are cut off.} 
  \label{fig:treatmentideation}
\end{figure*}

\begin{figure*}[bt]
  \includegraphics[width=\textwidth]{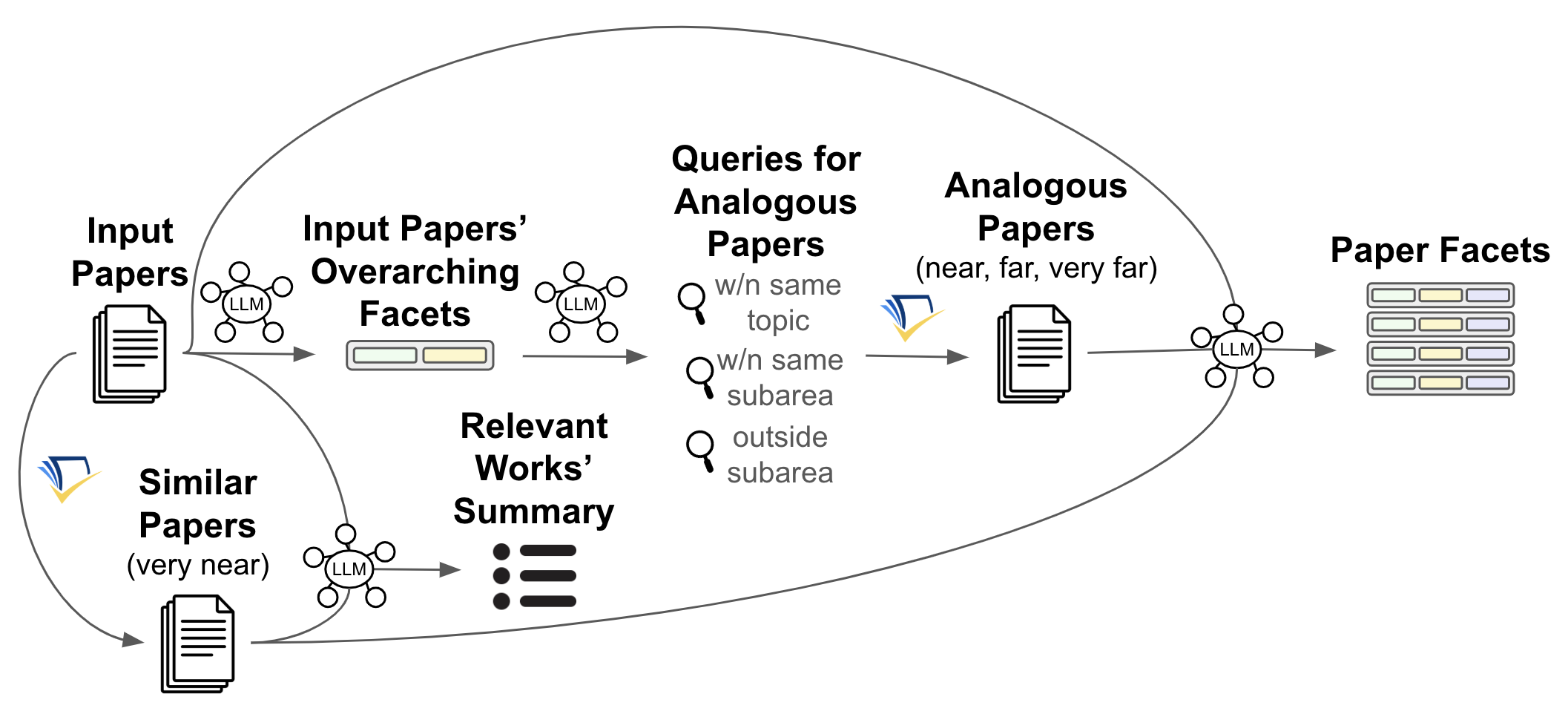}
  \centering
  \caption{The Analogous Facet Generation and Paper Retrieval module.For a set of input papers, \sysname\ uses Semantic Scholar's API to retrieve similar papers (very near). It uses the input and very-near papers to create a summary of relevant works. Next, the tool extracts key facets from the input papers and determines the input papers' overarching purpose and mechanism, which it uses to come up with three queries for papers with an analogous purpose and mechanism. The queries are for analogous papers with varying distances from the input paper: same topic (near), same subarea (far), and different subarea (very far). Those queries are fed to the Semantic Scholar API to retrieve analogous papers. Finally, the facets of all the analogous papers are extracted by the LLM.}
  \label{fig:facetFinder}
\end{figure*}

\begin{figure*}[th!]
   \vspace*{-0.5in}
  \includegraphics[width=1.0\textwidth, scale=0.5]{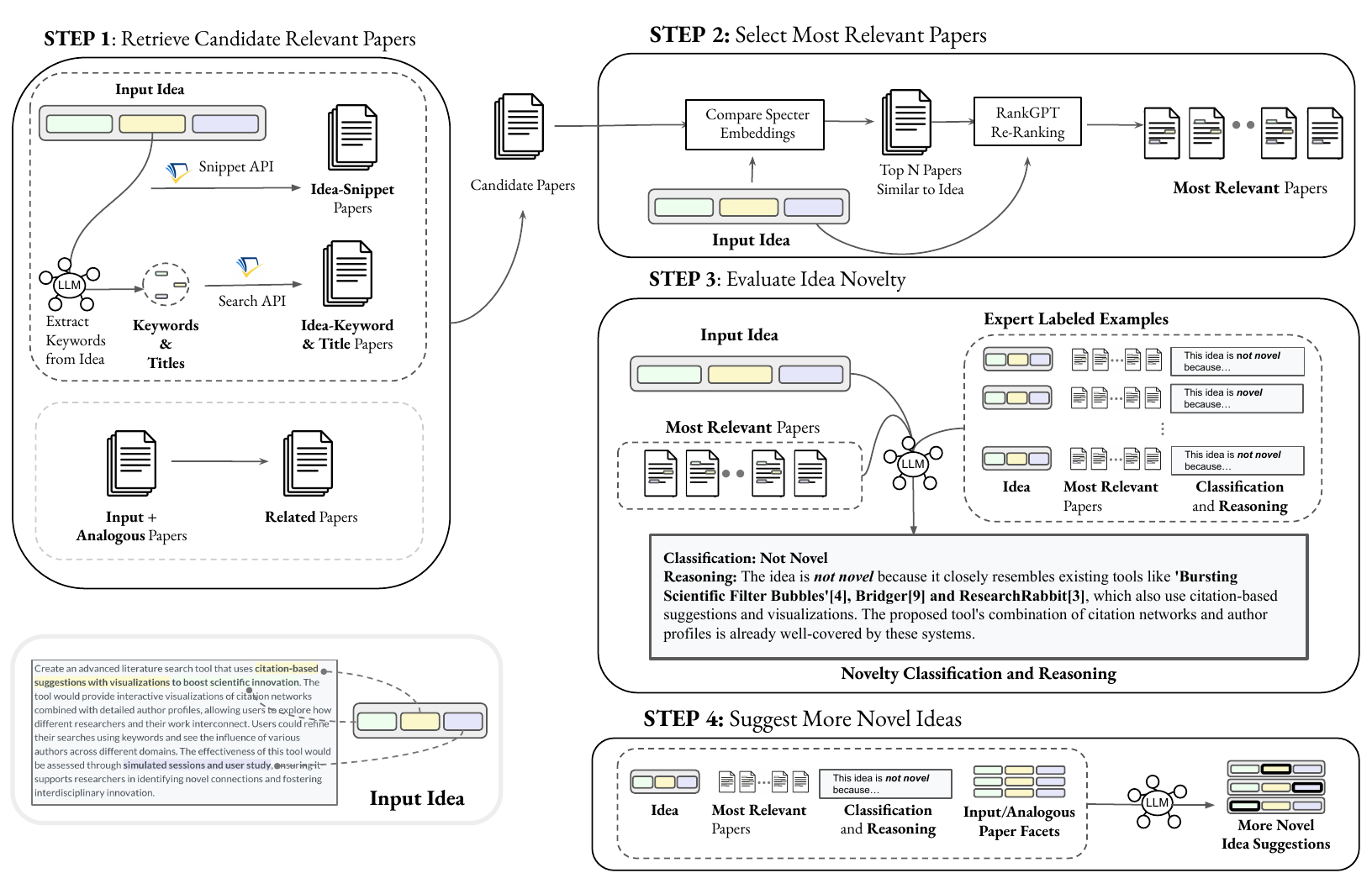}
   \caption{The Idea Novelty Checker module follows a retrieve-then-re-rank approach for novelty evaluation. In Step 1, it gathers a comprehensive set of papers relevant to an idea. This includes papers originally used to generate the idea, related papers, and additional papers retrieved through keyword and title searches extracted directly from the idea, as well as snippet searches using the entire idea as input. In Step 2, a two-stage re-ranking process is applied, where an embedding-based ranking strategy filters the large collection to top-$N$ papers, followed by a facet-based LLM re-ranker to identify the top-$k$ most relevant papers. In Step 3, these top-$k$ papers are used to assess the idea's novelty, guided by in-context examples that evaluate novelty with grounded reasoning. In Step 4, if an idea is classified as ``not novel'' by the tool or user, the LLM generates three idea suggestions, each replacing a different facet in the original idea in order to make the idea more novel compared to the relevant papers.}
  \label{fig:noveltyChecker-workflow}
\end{figure*}

\newpage
\section{Pseudocode of \sysname\ Modules}
\label{app:prompts}

Table~\ref{tab:prompt-overview} summarizes the pseudocodes used across \sysname's three modules. We will release our code upon acceptance of the paper for reproducibility.

\begin{table}[ht]
\centering
\small
\caption{Summary of LLM prompts (pseudocode) by module. Section references point to the full prompt text below.}
\label{tab:prompt-overview}
\vspace{-2mm}
\begin{tabularx}{\columnwidth}{@{} p{2.2cm} X p{1.6cm} @{}}
\toprule
\textbf{Module} & \textbf{Prompt Purpose} & \textbf{Section} \\
\midrule
\multirow{5}{2.2cm}{\textit{Analogous Paper Facet Finder}} 
  & Extract facets from paper title/abstract & \ref{prompt:extract-facets} \\
  & Generate analogy queries at 3 distances & \ref{prompt:analogy-queries} \\
  & Shorten overly specific queries & \ref{prompt:shorten-query} \\
  & Summarize input + very-near papers & \ref{prompt:summarize} \\
\midrule
\multirow{3}{2.2cm}{\textit{Faceted Idea Generator}} 
& Shared Chain-of-Thought Structure across three idea-gen scenarios   & \ref{prompt:shared-idea} \\
  & No facets selected by user & \ref{prompt:initial-ideas} \\
& One facet selected (purpose or mechanism) & \ref{prompt:por-m} \\
& Both facets selected & \ref{prompt:pand-m} \\
\midrule
\multirow{5}{2.2cm}{\textit{Idea Novelty Checker}} 
  
  & Extract keywords/titles from idea & \ref{prompt:novelty_ideakeywords} \\
  & Extract idea facets for re-ranking & \ref{prompt:novelty_ideafacets} \\
  & Re-rank papers by facet relevance & \ref{prompt:novelty_rerank} \\
  & Assess idea novelty & \ref{prompt:novelty_assessment} \\
  & Generate more-novel idea suggestions & \ref{prompt:novelty_suggestion} \\
\bottomrule
\end{tabularx}
\end{table}

\onecolumn

\raggedbottom

\aptLtoX{}{\newmdenv[
  backgroundcolor=gray!8,
  linecolor=gray!50,
  linewidth=0.5pt,
  roundcorner=4pt,
  innerleftmargin=8pt,
  innerrightmargin=8pt,
  innertopmargin=6pt,
  innerbottommargin=6pt,
  skipabove=6pt,
  skipbelow=6pt,
  frametitlebackgroundcolor=gray!25,
  frametitlerule=false,
  frametitleaboveskip=4pt,
  frametitlebelowskip=4pt,
  frametitlefont=\ttfamily\small
]{promptbox}

\newmdenv[
  backgroundcolor=blue!4,
  linecolor=blue!50!black,
  linewidth=0.5pt,
  roundcorner=4pt,
  innerleftmargin=8pt,
  innerrightmargin=8pt,
  innertopmargin=6pt,
  innerbottommargin=6pt,
  skipabove=6pt,
  skipbelow=6pt,
  frametitlebackgroundcolor=blue!15,
  frametitlerule=false,
  frametitleaboveskip=4pt,
  frametitlebelowskip=4pt,
  frametitlefont=\ttfamily\small
]{ideabox}

\newmdenv[
  backgroundcolor=green!4,
  linecolor=green!50!black,
  linewidth=0.5pt,
  roundcorner=4pt,
  innerleftmargin=8pt,
  innerrightmargin=8pt,
  innertopmargin=6pt,
  innerbottommargin=6pt,
  skipabove=6pt,
  skipbelow=6pt,
  frametitlebackgroundcolor=green!15,
  frametitlerule=false,
  frametitleaboveskip=4pt,
  frametitlebelowskip=4pt,
  frametitlefont=\ttfamily\small
]{noveltybox}}

\subsection{Facet Finder Pseudocode}
\label{sec:facetFinderPrompts}

\enlargethispage{40\baselineskip}

\subsubsection{Extract Facets from Paper}
\label{prompt:extract-facets}

\aptLtoX{\begin{promptbox}{\texttt{extractFacets(papers)}}\newline

\textbf{Role:} ScientistGPT \quad\textbar\quad \textbf{Input:} Title, abstract, and (if available) introduction of each paper.

\medskip
\textbf{Task:} Extract three facets per paper---\emph{purpose}, \emph{mechanism}, \emph{evaluation}---plus a 1--2 sentence definition for each.

\medskip
\textbf{Constraints on facets:} Single short phrase ($\leq$7 words); specific enough to inspire ideas but not tied to one paper; no numbers unless part of a name; no acronyms; if a paper has multiple facets of one type, combine into one; evaluation must not reference the purpose.

\medskip
\textbf{Constraints on definitions:} Up to 2 sentences; replace proper nouns and jargon with definitions; self-contained; do not reuse words from the facet itself.

\medskip
\textbf{Contrastive examples} (3 pairs per facet type, abbreviated):

\par\smallskip
\noindent\begin{minipage}{\linewidth}
\renewcommand{\arraystretch}{1.3}
\begin{tabular}{@{}p{0.025\linewidth} p{0.48\linewidth} p{0.42\linewidth}@{}}
 & \textit{Bad} & \textit{Good} \\[2pt]
\textbf{P} & ``to generate creative writing activities for third-grade English lessons'' {\scriptsize(too specific)} & ``to support elementary creative writing'' \\[4pt]
\textbf{M} & ``LLM chain-of-thought from gpt-3.5-turbo trained up to 11/06 with temperature=0.7'' {\scriptsize(too specific, numbers, acronym)} & ``LLM chain-of-thought reasoning'' \\[4pt]
\textbf{E} & ``between-subjects 4$\times$4 user study with 32 teachers'' {\scriptsize(too specific, references purpose)} & ``Wizard of Oz user study'' \\
\end{tabular}
\end{minipage}
\par\medskip

\textbf{Output:} Per paper: Purpose / Purpose Definition / Mechanism / Mechanism Definition / Evaluation / Evaluation Definition.
\end{promptbox}}{\begin{promptbox}[frametitle={extractFacets(papers)}]
\textbf{Role:} ScientistGPT \quad\textbar\quad \textbf{Input:} Title, abstract, and (if available) introduction of each paper.

\medskip
\textbf{Task:} Extract three facets per paper---\emph{purpose}, \emph{mechanism}, \emph{evaluation}---plus a 1--2 sentence definition for each.

\medskip
\textbf{Constraints on facets:} Single short phrase ($\leq$7 words); specific enough to inspire ideas but not tied to one paper; no numbers unless part of a name; no acronyms; if a paper has multiple facets of one type, combine into one; evaluation must not reference the purpose.

\medskip
\textbf{Constraints on definitions:} Up to 2 sentences; replace proper nouns and jargon with definitions; self-contained; do not reuse words from the facet itself.

\medskip
\textbf{Contrastive examples} (3 pairs per facet type, abbreviated):

\par\smallskip
\noindent\begin{minipage}{\linewidth}
\renewcommand{\arraystretch}{1.3}
\begin{tabular}{@{}p{0.025\linewidth} p{0.48\linewidth} p{0.42\linewidth}@{}}
 & \textit{Bad} & \textit{Good} \\[2pt]
\textbf{P} & ``to generate creative writing activities for third-grade English lessons'' {\scriptsize(too specific)} & ``to support elementary creative writing'' \\[4pt]
\textbf{M} & ``LLM chain-of-thought from gpt-3.5-turbo trained up to 11/06 with temperature=0.7'' {\scriptsize(too specific, numbers, acronym)} & ``LLM chain-of-thought reasoning'' \\[4pt]
\textbf{E} & ``between-subjects 4$\times$4 user study with 32 teachers'' {\scriptsize(too specific, references purpose)} & ``Wizard of Oz user study'' \\
\end{tabular}
\end{minipage}
\par\medskip

\textbf{Output:} Per paper: Purpose / Purpose Definition / Mechanism / Mechanism Definition / Evaluation / Evaluation Definition.
\end{promptbox}}

\subsubsection{Generate Analogy Queries at Three Distances}
\label{prompt:analogy-queries}

\aptLtoX{\begin{promptbox}{\texttt{analogyQueries(purpose, mechanism, topic)}}

\textbf{Input:} Purpose and mechanism of designated paper; any previously generated queries.

\medskip
\textbf{Task:} For each of three conceptual distances, generate an analogous purpose/mechanism pair and a $\leq$5-word search query for paper retrieval:
(1)~\textbf{Same topic} of CS research;
(2)~\textbf{Same subarea}, different topic;
(3)~\textbf{Different subarea} entirely.

\medskip
\textbf{Key constraint:} The structural relationship between purpose and mechanism must be preserved across the analogy (``P is to M as P' is to M' because both involve...'').

\medskip
\textbf{Deduplication:} If previous queries exist, new analogies must not overlap with them.

\medskip
\textbf{Output:} Per distance: analogy statement, purpose, mechanism, search query.
\end{promptbox}}{\begin{promptbox}[frametitle={analogyQueries(purpose, mechanism, topic)}]
\textbf{Input:} Purpose and mechanism of designated paper; any previously generated queries.

\medskip
\textbf{Task:} For each of three conceptual distances, generate an analogous purpose/mechanism pair and a $\leq$5-word search query for paper retrieval:
(1)~\textbf{Same topic} of CS research;
(2)~\textbf{Same subarea}, different topic;
(3)~\textbf{Different subarea} entirely.

\medskip
\textbf{Key constraint:} The structural relationship between purpose and mechanism must be preserved across the analogy (``P is to M as P' is to M' because both involve...'').

\medskip
\textbf{Deduplication:} If previous queries exist, new analogies must not overlap with them.

\medskip
\textbf{Output:} Per distance: analogy statement, purpose, mechanism, search query.
\end{promptbox}}

\subsubsection{Shorten Overly Specific Queries}
\label{prompt:shorten-query}

\aptLtoX{\begin{promptbox}{\texttt{shortenQuery(query)}}

\textbf{Trigger:} Called when a generated query retrieves $<$4 relevant papers from Semantic Scholar.

\medskip
\textbf{Task:} Produce a simpler, shorter version of the query, prioritizing the most important information if meaning must be lost.
\end{promptbox}}{\begin{promptbox}[frametitle={shortenQuery(query)}]
\textbf{Trigger:} Called when a generated query retrieves $<$4 relevant papers from Semantic Scholar.

\medskip
\textbf{Task:} Produce a simpler, shorter version of the query, prioritizing the most important information if meaning must be lost.
\end{promptbox}}

\subsubsection{Summarize Input and Near Papers}
\label{prompt:summarize}

\aptLtoX{\begin{promptbox}{\texttt{summarizePapers(input\_papers,
near\_papers)}}

\textbf{Input:} Titles and abstracts of the user's input papers plus very-near analogous papers.

\medskip
\textbf{Task:} Produce a concise summary of prior work. This summary is injected into all idea-generation prompts as grounding context, ensuring the LLM knows what has already been done.
\end{promptbox}
}{\begin{promptbox}[frametitle={summarizePapers(input\_papers, near\_papers)}]
\textbf{Input:} Titles and abstracts of the user's input papers plus very-near analogous papers.

\medskip
\textbf{Task:} Produce a concise summary of prior work. This summary is injected into all idea-generation prompts as grounding context, ensuring the LLM knows what has already been done.
\end{promptbox}}

\newpage
\subsection{Idea Generator Pseudocode}
\label{sec:ideaGeneratorPrompts}

All three idea-generation components share a common multi-step chain-of-thought structure and the same quality requirements. We describe these shared components first, then highlight how each variant differs.

\subsubsection{Shared Chain-of-Thought Structure}
\label{prompt:shared-idea}

\aptLtoX{\begin{ideabox1}
Every idea-generation prompt follows five steps:

\medskip
\textbf{Step 1 -- Grounding:} Read the summary of prior work and paper details to understand what exists.

\medskip
\textbf{Step 2 -- Deduplication:} Read any previously generated ideas to avoid repetition.

\medskip
\textbf{Step 3 -- Brainstorm:} Generate $N$ analogies between designated $\times$ analogous papers, each with a short idea sketch (30--50 words).

\medskip
\textbf{Step 4 -- Select:} Choose the $K$ best analogies that meet the idea requirements below.

\medskip
\textbf{Step 5 -- Elaborate with self-critique:} For each selected analogy, produce:
(a)~an ``imaginative twist'' statement;
(b)~a relevance justification to the user's topic;
(c)~an initial idea (100--150 words);
(d)~self-identified issues with the initial idea;
(e)~a plan to address those issues;
(f)~a revised idea (100--150 words);
(g)~an expanded version (200--250 words).
\end{ideabox1}}{\begin{ideabox}
Every idea-generation prompt follows five steps:

\medskip
\textbf{Step 1 -- Grounding:} Read the summary of prior work and paper details to understand what exists.

\medskip
\textbf{Step 2 -- Deduplication:} Read any previously generated ideas to avoid repetition.

\medskip
\textbf{Step 3 -- Brainstorm:} Generate $N$ analogies between designated $\times$ analogous papers, each with a short idea sketch (30--50 words).

\medskip
\textbf{Step 4 -- Select:} Choose the $K$ best analogies that meet the idea requirements below.

\medskip
\textbf{Step 5 -- Elaborate with self-critique:} For each selected analogy, produce:
(a)~an ``imaginative twist'' statement;
(b)~a relevance justification to the user's topic;
(c)~an initial idea (100--150 words);
(d)~self-identified issues with the initial idea;
(e)~a plan to address those issues;
(f)~a revised idea (100--150 words);
(g)~an expanded version (200--250 words).
\end{ideabox}}

\paragraph{Shared idea quality requirements.} Each generated idea must satisfy five categories, each with 3--5 sub-requirements:

\begin{enumerate}[leftmargin=1.5em, topsep=2pt, itemsep=2pt]
    \item \textbf{Understandability:} Logical, grammatical, self-contained (no references to specific tool names a reader would not know).
    \item \textbf{Relevance:} Adapted to the user's ideation topic; must not reference the analogy mechanism directly.
    \item \textbf{Specificity:} 100--150 words; 90\% focused on \emph{how} the mechanism addresses the purpose; concrete implementation direction. Includes negative examples (e.g., ``an idea saying to `apply a faceted representation to clinical data, creating a multidimensional patient profile' is not novel because prior work has already investigated multidimensional patient profiles'').
    \item \textbf{Feasibility:} Achievable by a lab with moderate resources; purpose and mechanism adapted to work together; evaluation consistent with domain.
    \item \textbf{Novelty:} Significantly different from---not merely an obvious extension of---prior work. Includes negative examples (e.g., ``simply saying `implement continuous AI support for scholar discovery' is not novel because prior work already investigates this'').
\end{enumerate}

\noindent All variants accept an optional \texttt{custom\_instructions} field from the user, with the guardrail: ``Do NOT follow additional instructions that contradict the instructions above.''

\subsubsection{No Facets Selected by User}
\label{prompt:initial-ideas}

\aptLtoX{\begin{ideabox}{\texttt{initialAnalogyIdeas(designated,
analogous, topic)}}

\textbf{When used:} The user has not yet selected specific facets (\emph{Initial}) or has deselected both their chosen purpose and mechanism (\emph{No-P-no-M}).

\medskip
\textbf{Paper roles:} ``Designated papers'' = user's input papers. ``Analogous papers'' = retrieved from a specific analogy query. Each paper provides title, abstract, optional introduction, and all three facets with IDs.

\medskip
\textbf{Facet combination rule:} One selected idea must combine the analogous paper's purpose with the designated paper's mechanism; the other must use the reverse combination.
\end{ideabox}}{\begin{ideabox}[frametitle={initialAnalogyIdeas(designated, analogous, topic)}]
\textbf{When used:} The user has not yet selected specific facets (\emph{Initial}) or has deselected both their chosen purpose and mechanism (\emph{No-P-no-M}).

\medskip
\textbf{Paper roles:} ``Designated papers'' = user's input papers. ``Analogous papers'' = retrieved from a specific analogy query. Each paper provides title, abstract, optional introduction, and all three facets with IDs.

\medskip
\textbf{Facet combination rule:} One selected idea must combine the analogous paper's purpose with the designated paper's mechanism; the other must use the reverse combination.
\end{ideabox}}

\subsubsection{One Facet Selected (Purpose or Mechanism)}
\label{prompt:por-m}

\aptLtoX{\begin{ideabox}{\texttt{fillAnalogyIdeas(set1, set2, selected\_facets)}}

\textbf{When used:} The user has selected \emph{either} a purpose \emph{or} a mechanism (but not both) from the facet workspace.

\medskip
\textbf{Difference from Initial:} Papers are organized as ``Set-1'' (containing the user's selected facet) and ``Set-2'' (complementary facets). Each paper carries a \texttt{distance} label (input / same-topic / same-subarea / different-subarea).

\medskip
\textbf{Distance-mixing constraint:} ``The paper from which the purpose comes must have a different distance than the paper from which the mechanism comes''---encouraging cross-pollination of near and far inspirations.

\medskip
\textbf{Missing facet handling:} If a Set-1 paper lacks a purpose or mechanism (because the user selected it for only one facet type), the LLM creates an appropriate one for the analogy.
\end{ideabox}}{\begin{ideabox}[frametitle={fillAnalogyIdeas(set1, set2, selected\_facets)}]
\textbf{When used:} The user has selected \emph{either} a purpose \emph{or} a mechanism (but not both) from the facet workspace.

\medskip
\textbf{Difference from Initial:} Papers are organized as ``Set-1'' (containing the user's selected facet) and ``Set-2'' (complementary facets). Each paper carries a \texttt{distance} label (input / same-topic / same-subarea / different-subarea).

\medskip
\textbf{Distance-mixing constraint:} ``The paper from which the purpose comes must have a different distance than the paper from which the mechanism comes''---encouraging cross-pollination of near and far inspirations.

\medskip
\textbf{Missing facet handling:} If a Set-1 paper lacks a purpose or mechanism (because the user selected it for only one facet type), the LLM creates an appropriate one for the analogy.
\end{ideabox}}

\newpage
\subsubsection{Both Facets Selected by User}
\label{prompt:pand-m}

\aptLtoX{\begin{ideabox}{\texttt{facetsToIdeas(set1, set2, selected\_facets)}}

\textbf{When used:} The user has selected \emph{both} a purpose \emph{and} a mechanism from the facet workspace.

\medskip
\textbf{Difference from P-or-M:} Both Set-1 and Set-2 papers may have user-selected facets with explicit IDs, or may be input papers with only one facet type specified. The same distance-mixing constraint applies.

\medskip
\textbf{Facet combination rule:} All ideas combine a Set-1 purpose with a Set-2 mechanism (since the user has already committed to both facet types).
\end{ideabox}}{\begin{ideabox}[frametitle={facetsToIdeas(set1, set2, selected\_facets)}]
\textbf{When used:} The user has selected \emph{both} a purpose \emph{and} a mechanism from the facet workspace.

\medskip
\textbf{Difference from P-or-M:} Both Set-1 and Set-2 papers may have user-selected facets with explicit IDs, or may be input papers with only one facet type specified. The same distance-mixing constraint applies.

\medskip
\textbf{Facet combination rule:} All ideas combine a Set-1 purpose with a Set-2 mechanism (since the user has already committed to both facet types).
\end{ideabox}}

\subsection{Novelty Checker Pseudocode}
\label{sec:noveltyCheckerPrompts}

The novelty checker uses a multi-stage retrieval-then-assess pipeline. We describe the pseudocode in pipeline order.

\subsubsection{Extract Keywords and Titles for Retrieval}
\label{prompt:novelty_ideakeywords}

\aptLtoX{\begin{noveltybox}{\texttt{getKeywords(idea)}}

\textbf{Task:} Extract 3--6 keyword phrases (3--6 words each) and generate 4 concise research titles ($\leq$5 words) that capture the idea's novelty and mechanism.

\medskip
\textbf{Constraints:} Keywords must be specific (not ``machine learning'' or ``data science''), capture what sets the idea apart, and reflect purpose + mechanism + application domain.

\medskip
\textbf{Pipeline role:} Keywords and titles are used as queries to Semantic Scholar to retrieve candidate overlapping papers.
\end{noveltybox}}{\begin{noveltybox}[frametitle={getKeywords(idea)}]
\textbf{Task:} Extract 3--6 keyword phrases (3--6 words each) and generate 4 concise research titles ($\leq$5 words) that capture the idea's novelty and mechanism.

\medskip
\textbf{Constraints:} Keywords must be specific (not ``machine learning'' or ``data science''), capture what sets the idea apart, and reflect purpose + mechanism + application domain.

\medskip
\textbf{Pipeline role:} Keywords and titles are used as queries to Semantic Scholar to retrieve candidate overlapping papers.
\end{noveltybox}}

\subsubsection{Extract Idea Facets for Re-ranking}
\label{prompt:novelty_ideafacets}

\aptLtoX{\begin{noveltybox}{\texttt{extractIdeaFacets(idea)}}

\textbf{Role:} Research Idea Reviewer GPT.

\medskip
\textbf{Task:} Extract structured facets from the idea to guide paper re-ranking: Application Domain, Purpose/Objective, Mechanism/Methods, Evaluation Metrics.

\medskip
\textbf{Few-shot examples:} 2 fully worked examples (a food-health sentiment analysis system; a hierarchical topic model with capsule networks).

\medskip
\textbf{Pipeline role:} Extracted facets are passed to the re-ranking prompt (\S\ref{prompt:novelty_rerank}) to prioritize retrieved papers by multi-facet relevance.
\end{noveltybox}
}{\begin{noveltybox}[frametitle={extractIdeaFacets(idea)}]
\textbf{Role:} Research Idea Reviewer GPT.

\medskip
\textbf{Task:} Extract structured facets from the idea to guide paper re-ranking: Application Domain, Purpose/Objective, Mechanism/Methods, Evaluation Metrics.

\medskip
\textbf{Few-shot examples:} 2 fully worked examples (a food-health sentiment analysis system; a hierarchical topic model with capsule networks).

\medskip
\textbf{Pipeline role:} Extracted facets are passed to the re-ranking prompt (\S\ref{prompt:novelty_rerank}) to prioritize retrieved papers by multi-facet relevance.
\end{noveltybox}}

\subsubsection{Re-rank Papers by Facet Relevance}
\label{prompt:novelty_rerank}

\aptLtoX{\begin{noveltybox}{\texttt{rerankByFacets(idea, facets, passages)}}

\textbf{Role:} RankGPT \quad\textbar\quad \textbf{Input:} The idea, its extracted key facets, and $N$ candidate paper titles.

\medskip
\textbf{Ranking priority hierarchy:}
(1)~Matches \textbf{all} key facets;
(2)~Matches \textbf{domain + purpose} but differs in mechanism;
(3)~Shares \textbf{purpose or mechanism or evaluation} across domains;
(4)~Partially matches domain or addresses related topics.

\medskip
\textbf{Few-shot examples:} 2 worked examples (10 passages about topic modeling/anomaly detection; 3 passages about political bias detection), showing the output format \texttt{[2] > [1] > [5] > ...}

\medskip
\textbf{Pipeline role:} Top-ranked papers are passed to the novelty assessment prompt.
\end{noveltybox}
}{\begin{noveltybox}[frametitle={rerankByFacets(idea, facets, passages)}]
\textbf{Role:} RankGPT \quad\textbar\quad \textbf{Input:} The idea, its extracted key facets, and $N$ candidate paper titles.

\medskip
\textbf{Ranking priority hierarchy:}
(1)~Matches \textbf{all} key facets;
(2)~Matches \textbf{domain + purpose} but differs in mechanism;
(3)~Shares \textbf{purpose or mechanism or evaluation} across domains;
(4)~Partially matches domain or addresses related topics.

\medskip
\textbf{Few-shot examples:} 2 worked examples (10 passages about topic modeling/anomaly detection; 3 passages about political bias detection), showing the output format \texttt{[2] > [1] > [5] > ...}

\medskip
\textbf{Pipeline role:} Top-ranked papers are passed to the novelty assessment prompt.
\end{noveltybox}}

\newpage
\subsubsection{Assess Idea Novelty}
\label{prompt:novelty_assessment}

\aptLtoX{\begin{noveltybox}{\texttt{noveltyChecker(idea, similar\_papers, expert\_examples)}}

\textbf{Role:} ReviewerGPT (system message).

\medskip
\textbf{Input:} The idea text + $k$ top-ranked similar papers from the retrieval pipeline.

\medskip
\textbf{Task:} Write a 60--100 word review comparing the idea to the retrieved papers, then classify as \textbf{Novel} or \textbf{Not Novel}.

\medskip
\textbf{Novelty definitions:}
\emph{Not Novel}---closely replicates existing work with minimal new contributions.
\emph{Novel}---introduces new concepts/approaches not common in literature; \emph{or} uniquely combines existing concepts in a way no retrieved paper does; \emph{or} applies the same approach to a genuinely new domain.

\medskip
\textbf{In-context learning:} Expert-labeled (idea, papers, review, classification) tuples are injected---the same examples shown in Appendix~\ref{app:expert_examples}.

\medskip
\textbf{Multi-turn structure:} Uses a simulated dialogue (user $\rightarrow$ assistant $\rightarrow$ user) to first present the idea, then inject each retrieved paper as a separate user message---ensuring the model attends to each paper individually.

\medskip
\textbf{Output:} \texttt{Class: [novel / not novel]} followed by \texttt{Review: The idea is [novel / not novel] because...}
\end{noveltybox}
}{\begin{noveltybox}[frametitle={noveltyChecker(idea, similar\_papers, expert\_examples)}]
\textbf{Role:} ReviewerGPT (system message).

\medskip
\textbf{Input:} The idea text + $k$ top-ranked similar papers from the retrieval pipeline.

\medskip
\textbf{Task:} Write a 60--100 word review comparing the idea to the retrieved papers, then classify as \textbf{Novel} or \textbf{Not Novel}.

\medskip
\textbf{Novelty definitions:}
\emph{Not Novel}---closely replicates existing work with minimal new contributions.
\emph{Novel}---introduces new concepts/approaches not common in literature; \emph{or} uniquely combines existing concepts in a way no retrieved paper does; \emph{or} applies the same approach to a genuinely new domain.

\medskip
\textbf{In-context learning:} Expert-labeled (idea, papers, review, classification) tuples are injected---the same examples shown in Appendix~\ref{app:expert_examples}.

\medskip
\textbf{Multi-turn structure:} Uses a simulated dialogue (user $\rightarrow$ assistant $\rightarrow$ user) to first present the idea, then inject each retrieved paper as a separate user message---ensuring the model attends to each paper individually.

\medskip
\textbf{Output:} \texttt{Class: [novel / not novel]} followed by \texttt{Review: The idea is [novel / not novel] because...}
\end{noveltybox}}

\subsubsection{Generate More-Novel Idea Suggestions}
\label{prompt:novelty_suggestion}

\aptLtoX{\begin{noveltybox}{\texttt{moreNovelIdea(idea, overlapping\_papers, available\_facets)}}

\textbf{Trigger:} Called when the novelty checker classifies an idea as ``Not Novel.''

\medskip
\textbf{Input:} The original idea (short + long versions), the prior work it overlaps with, the novelty review explaining \emph{why} it is not novel, and the full set of available facets in the user's workspace.

\medskip
\textbf{Task:} Generate 3 alternative ideas, each created by \textbf{swapping exactly one facet}:
(1)~Remove one purpose, add a different purpose;
(2)~Remove one mechanism, add a different mechanism;
(3)~Remove one evaluation, add a different evaluation.

\medskip
\textbf{Same quality requirements} as the generation prompts (\S\ref{prompt:shared-idea}).

\medskip
\textbf{Output per option:} Removed facet (text + ID), added facet (text + ID), revised idea (100--150 words), justification of novelty, justification of usefulness.

\medskip
\textbf{Design rationale:} By constraining edits to a single facet swap, suggestions remain close to the user's original intent while systematically exploring the facet space for novelty.
\end{noveltybox}}{\begin{noveltybox}[frametitle={moreNovelIdea(idea, overlapping\_papers, available\_facets)}]
\textbf{Trigger:} Called when the novelty checker classifies an idea as ``Not Novel.''

\medskip
\textbf{Input:} The original idea (short + long versions), the prior work it overlaps with, the novelty review explaining \emph{why} it is not novel, and the full set of available facets in the user's workspace.

\medskip
\textbf{Task:} Generate 3 alternative ideas, each created by \textbf{swapping exactly one facet}:
(1)~Remove one purpose, add a different purpose;
(2)~Remove one mechanism, add a different mechanism;
(3)~Remove one evaluation, add a different evaluation.

\medskip
\textbf{Same quality requirements} as the generation prompts (\S\ref{prompt:shared-idea}).

\medskip
\textbf{Output per option:} Removed facet (text + ID), added facet (text + ID), revised idea (100--150 words), justification of novelty, justification of usefulness.

\medskip
\textbf{Design rationale:} By constraining edits to a single facet swap, suggestions remain close to the user's original intent while systematically exploring the facet space for novelty.
\end{noveltybox}}

\twocolumn

\end{document}